\documentclass[traditabstract]{aa}
\usepackage{graphicx}
\usepackage{txfonts}
\usepackage{todonotes}
\usepackage[normalem]{ulem}
\usepackage{amstext}
\usepackage[breaklinks=true]{hyperref}
\usepackage{xspace}
\usepackage{lscape}
\usepackage{mathrsfs}

\usepackage{listings}
\makeatletter
\renewcommand*\maketitle{%
  \thispagestyle{firstpage}
\begingroup
    \if@wideboxfn
    \setlength\bibindent{1.4\parindent}
    \else
    \setlength\bibindent{\parindent}
    \fi
    \renewcommand*\thefootnote{\@fnsymbol\c@footnote}%
    \renewcommand\@makefntext[1]{%
    \ifaa@longfn\hsize\textwidth\fi
    \noindent
    \hb@xt@\bibindent{\hss\@makefnmark\enspace}##1}
  \ifaa@twocolumn
  \begingroup
    \begin{aa@strip}
          \aa@maketitle
    \end{aa@strip}
    \@thanks            
  \endgroup
  \else
    \begingroup
      \let\thanks\footnote
      \aa@maketitle
    \endgroup
  \fi
\endgroup
  \setcounter{footnote}{0}%
}
\makeatother

\definecolor{dkgreen}{rgb}{0,0.6,0}
\definecolor{gray}{rgb}{0.5,0.5,0.5}
\definecolor{mauve}{rgb}{0.58,0,0.82}
\newcommand{\orcit}[1]{\protect\href{https://orcid.org/#1}{\protect\includegraphics[width=8pt]{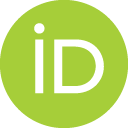}}}

\newcommand{\Gaia}{\textit{Gaia}\xspace}

\def\ltsima{$\, \buildrel < \over \sim \,$}
\def\simlt{\lower.5ex\hbox{\ltsima}}
\def\gtsima{$\, \buildrel > \over \sim \,$}
\def\simgt{\lower.5ex\hbox{\gtsima}}

\newcommand{\mtip}  {m_0^{TRGB}}
\newcommand{\mdown} {m_{\rm down}}
\newcommand{\mup}   {m_{\rm up}}
\newcommand{\LL}    {\mathcal{L}}
\newcommand{\LLs}   {\mathcal{L}_{\sigma}}
\newcommand{\dd}    {\text{d}}
\newcommand{\erf}   {\text{erf}}
\newcommand{\LLRGB}  {\mathcal{L}_{\rm RGB}}
\newcommand{\mi}    {m_i}

\newcommand{\sigtip}{\sigma_{TRGB}}
\begin{document}

\title{The RGB tip in the SDSS, PS1, JWST, NGRST and Euclid photometric systems
} 
\subtitle{Calibration in optical passbands using Gaia DR3 synthetic photometry}
\authorrunning{M. Bellazzini and R. Pascale}
\titlerunning{
The RGB tip in the SDSS, PS1, JWST, NGRST and Euclid photometric systems}

\author{
M.~                    Bellazzini\orcit{0000-0001-8200-810X}\inst{1}
\and         
R.~                       Pascale\orcit{}\inst{1}}
\institute{
INAF - Osservatorio di Astrofisica e Scienza dello Spazio di Bologna, via Piero Gobetti 93/3, 40129 Bologna, Italy
}




\date{Accepted for publication on August 8, 2024}

\abstract{We use synthetic photometry from Gaia DR3 BP and RP spectra for a large selected sample of stars in the Large Magellanic Cloud (LMC) and Small Magellanic Cloud (SMC) to derive the magnitude of the Red Giant Branch (RGB) tip for these two galaxies in several passbands in various widely used optical photometric systems, including those of space missions that have not yet started operations. 
The RGB tip is estimated by fitting a well-motivated model to the RGB luminosity function (LF) within a fully Bayesian framework, allowing for a proper representation of the uncertainties of all the involved parameters and their correlations. 
Adopting the best available distance and interstellar extinction estimates  we provide a calibration of the RGB tip as a standard candle for the following passbands: Johnson-Kron-Cousins I (mainly used for validation purposes), Hubble Space Telescope F814W, Sloan Digital Sky Survey i and z, PanSTARRS~1 y, James Webb Space Telescope F090W, Nancy Grace Roman Space Telescope Z087, and Euclid I$_E$, with an accuracy of a few per cent, depending on the case. The trend of the absolute magnitude of the tip as a function of colour in the different passbands, beyond the range spanned by the LMC and SMC, as well as its dependency on age, is explored by means of theoretical models.
These calibrations can be very helpful to obtain state-of-the-art RGB tip distance estimates to stellar systems in a very large range of distances directly from data in the natural photometric system of these surveys and/or missions, without recurring to photometric transformations. 
We make the photometric catalogues publicly available for calibrations in additional passbands or for different approaches in the estimate of the tip, as well as for stellar populations and stellar astrophysics studies that may take advantage of large and homogeneous datasets of stars with magnitudes in 22 different passbands.}


\maketitle

\section{Introduction}
\label{sec:intro}

The tip of the red giant branch (RGB) is a well-known and well-understood feature of the color-magnitude diagram (CMD) of stellar systems hosting significant populations of stars older than $\simeq 2$~Gyr \citep[see, e.g.,][and references therein]{salacas98, barker04, mbtip08, serenelli17, madore20, li23}. For stars in the mass range $\simeq 0.5-2.0~M_{\sun}$ the luminosity of RGB stars mainly depends on the mass of the degenerate He core ($M_c$) that is nearly constant at the violent onset of the core He-burning phase \citep[He flash,][]{serenelli17}. The sudden change in the evolutionary phase from shell H-burning (RGB) to core He-burning (red clump, RC, for the intermediate-age  stars, and horizontal branch, HB, for the oldest stars) induce a strong discontinuity at the bright end of the RGB LF, that is the actual observable we refer to with the term RGB tip \citep{lee93, madore95}. The constancy of $M_c$ at the tip and, consequently, the nearly constancy of the luminosity of the feature provide the key characteristics for a standard candle. In particular, the magnitude of the RGB tip in the reddest optical passbands, typically the Johnson-Kron-Cousins I band, has been shown to be very mildly { dependent} on metallicity (as well as on other stellar parameters), and nearly constant for [Fe/H]$\le -0.7$ \citep{dacosta90, barker04, mbtip08, serenelli17}, while it correlates linearly with metallicity in the near infrared passbands
\citep[see, e.g.,][and references therein]{mbtip04, valenti04, serenelli17}.

Introduced in its modern form in the early '90 \citep{lee93, madore95, sakai96} the use of the RGB tip as a standard candle has become a precious and widely adopted tool in the establishment of the local and cosmological distance scale \citep[see, e.g.,][and references therein]{maiz-apellaniz02, bellaz04, bellaz05, conn12, conn16, freedman19, anand21, scolnic23,madore23}. The behaviour of the observable and of the associated biases and uncertainties have been also extensively studied \citep[][]{madore95, salacas98, bellaz02,barker04,serenelli17, madore20, saltas22, madore23, wu23}.

While classical calibrations of the RGB tip were based on the RR Lyrae / HB distance scale \citep[see, e.g.,][]{lee93, rizzi07} or on theoretical models \citep{salacas98}, \citet{mbtip01} introduced a new calibration whose zero point was anchored to the semi-geometrical distance to the massive globular cluster $\omega$~Centauri obtained by \citet{omega01} from a double-lined detached eclipsing binary within the cluster, OGLE~17.
Two decades later, accurate distance estimates from eclipsing binaries have become available for two local galaxies that are classical pillars of the cosmological distance scale, the Large Magellanic Cloud \citep[LMC;][]{lmc19} and the Small Magellanic Cloud \citep[SMC;][]{smc20}. \citet[][H23 hereafter]{hoyt23} took advantage of these new distance measures, of the large sample of LMC and SMC photometry provided by the The Optical Gravitational Lensing Experiment (OGLE) project \citep{udalmc08,udasmc08} and of new reddening maps from the same project \citep{skowron21}, to derive, through a careful and insightful analysis, a new calibration of the RGB tip in the I band ensuring $\simeq 1\%$ consistency in distance  (see \citealt{soltis21,li22,li23,dixon23} for recent attempts of direct calibration of the standard candle using astrometry from the ESA-Gaia mission  \citealt{Prusti2016,edr3_general}).

In the present contribution we follow the path traced by H23 but using large samples of LMC and SMC stars with synthetic photometry from the Gaia-DR3 \citep{dr3_general} externally calibrated BP and RP (XP hereafter) spectra \citep{dr3_xp_fda,dr3_xp_paolo}. As illustrated in detail in \citet{dr3_dpacp93}, especially for the red passbands that are the most suitable for a safe use of the RGB tip as a standard candle, synthetic photometry accurate to 2-3\% over a large range of colours (spectral types) can be obtained from XP spectra for any wide passband enclosed in their spectral range. Moreover, \citet{dr3_dpacp93} used selected samples of photometric standard stars to enhance the accuracy of the XP synthetic photometry in a few widely used photometric systems, with a process thay call {\em standardisation}. In these cases photometric accuracy at a few millimag level is achieved. 

\citet{mbtip08} suggested, and, in some cases, showed by means of theoretical models, that other passbands sampling the stellar spectrum in the range 700~nm$\la \lambda \la$ 1100~nm may be effective in tracing the RGB tip with minimal dependency on metallicity and age, as for ${\rm I_{JKC}}$.
Taking advantage of the great flexibility allowed by XP synthetic photometry we directly estimate the magnitude of the RGB tip in several suitable passbands in different photometric systems, thus extending the calibration of the RGB tip as a standard candle beyond the JKC system, to other widely used systems as those of the Sloan Digital Sky Survey \citep[SDSS][]{Fukugita1996} or Pan-STARRS~1 \citep[PS1][]{Magnier2020}, as well as those of space missions that have recently started operations like the James Webb Space Telescope
(JWST\footnote{\url{https://webb.nasa.gov}}) and Euclid\footnote{\url{https://sci.esa.int/web/euclid}}, or are still to be launched in space, like the Nancy Grace Roman Space Telescope (NGSRT\footnote{\url{https://roman.gsfc.nasa.gov}}). In this way  we provide the tools to derive distance estimates from observations of the RGB of stellars systems over a very wide range of distances directly from the observations performed in the photometric systems of these surveys and space missions, without recurring to uncertain and possibly biased photometric transformations \citep[see][for discussion and references]{dr3_dpacp93}.


The plan of the paper is the following. In Sect.~\ref{sec:samples} we present the samples of SMC and LMC stars that we use to estimate the magnitude of the tip in different passbands and how we correct observed synthetic magnitudes of these stars for interstellar extinction.  In Sect.~\ref{sec:measure} we describe the results of our measures of the RGB tip and we provide the calibrations of the absolute magnitude of the tip in different passbands as a function of colour, extending the colour range of applicability with the help of theoretical models.
In Sect.~\ref{sec:algo} we introduce the bayesian method that we use to measure the tip, briefly discussing its robustness, quality and limitations. Finally, we summarise our results and draw our conclusions in Sect.~\ref{sec:conclu}.

In Appendix~\ref{app:dete} we display the diagrams of the actual detection of the tip in the various considered passbands. In Appendix~\ref{app:diff_hoyt} we discuss the possible reasons for the $\simeq 0.05$~mag difference in our estimates of the RGB tip of the LMC and SMC with respect to those obtained by H23, while in Appendix~\ref{app:cata} we briefly describe the LMC and SMC catalogues with XPSP in 22 passbands that we used in the present study and that we make publicly available. Finally, in Appendix~\ref{app:corner} we report on the parameters obtained from the RGB tip fitting procedure used, their uncertainties, and give further details on the tests we performed.

\section{The samples}
\label{sec:samples}

As a first step, we consider here only stars that are included in the \Gaia Synthetic Photometry Catalogue \citep[GSPC,][]{dr3_dpacp93}\footnote{The table {\tt gaiadr3.synthetic\_photometry}, in the Gaia Archive \url{https://gea.esac.esa.int/archive/}} to avoid stars with low photometric signal-to-noise (S/N) ratio in any of the GSPC passbands. Then, to remove sources with possibly flawed astrometry, we consider only stars with {\tt RUWE < 1.3} \citep{ruwe}. To minimise the effects of blendings and contamination in the photometry of individual stars we also require $|C^{\star}|<\sigma_{C^{\star}}(G)$, according to Eq.~18 of \citet{edr3_riello21}. 

Hence, we selected from the GSPC stars satisfying the above quality criteria and lying (a) within 10 degrees from the center of the LMC and within $2.0$~mas/yr from its mean proper motion (LMC sample), and (b) within 8 degrees from the center of the SMC and within $1.5$~mas/yr from its mean proper motion (SMC sample), where the galaxy centers and mean proper motions at the center have been taken from \citet{edr3_luri}. Finally, removing all stars having error on the parallax lower than 20\% (18242 and 3752 for the LMC and SMC samples, respectively) was a simple and very efficient mean to select out foreground Galactic stars. The resulting LMC sample contains 603311 stars, while the SMC sample contains 124578 stars\footnote{These catalogues are publicly available at \url{https://zenodo.org/records/10636449}. See Appendix~\ref{app:cata} for a detailed description of the content.}.

For each of these stars a value of the interstellar reddening E(B-V) was derived from the maps by \citet{skowron21}, when available, and from the maps by \citet{sfd98}, recalibrated by \citet{schlafly2011} in regions not reached by the former map. \citet{skowron21} reddening values are in the same scale as \citet{schlafly2011}; this has been verified for the stars in common in the outer regions of the galaxies where the comparison is meaningful \citep{sfd98}. To convert E(V-I) provided by \citet{skowron21} into E(B-V) we used the equation { E(B-V)=E(V-I)/1.399, from \citet[][their Table~6, see below]{schlafly2011}.}

\begin{figure*}[ht!]
\center{
\includegraphics[width=\columnwidth]{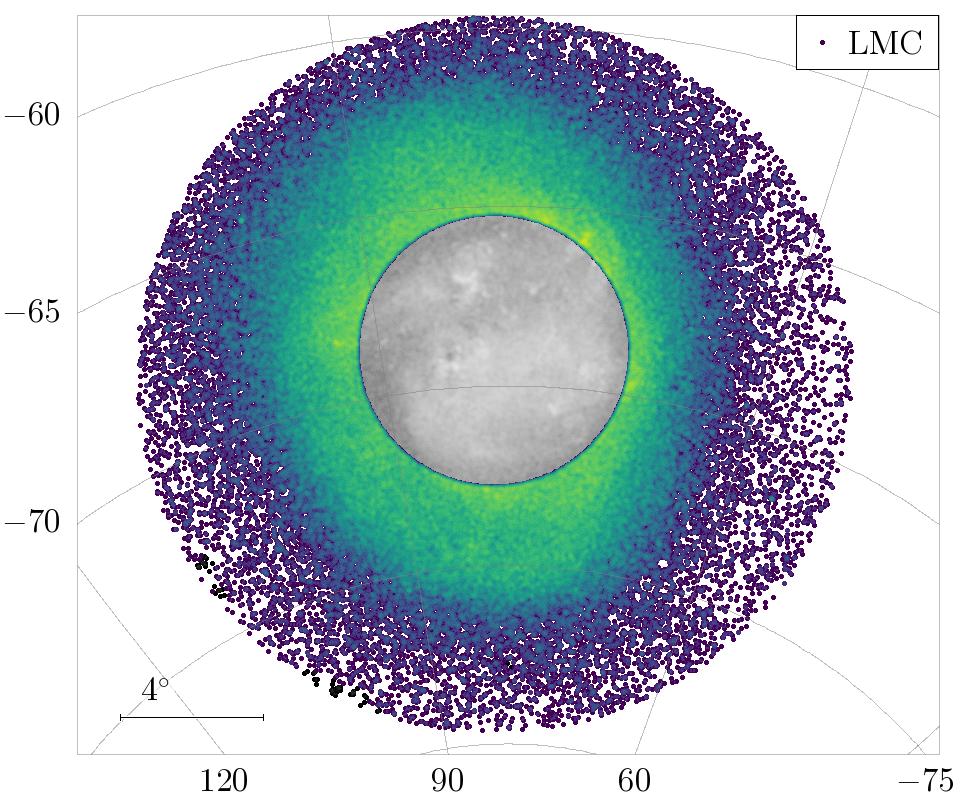}
\includegraphics[width=\columnwidth]{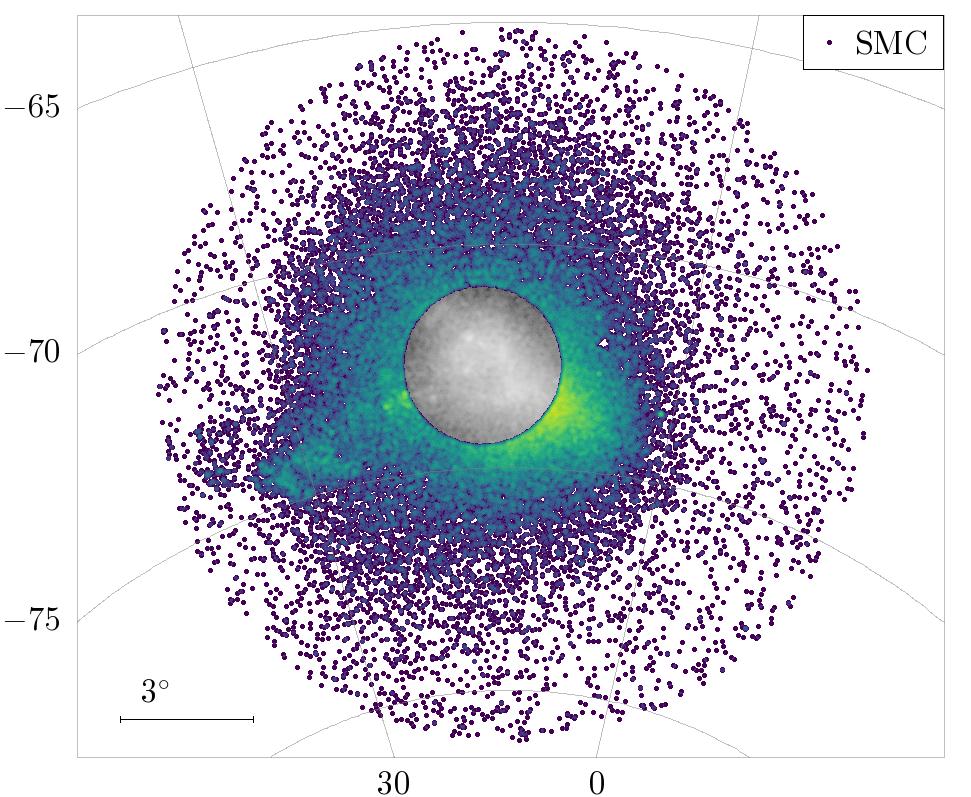}
}
\caption{Distribution in the sky of the stars in our LMC sample (left panel) and SMC sample (right panel). Stars are colour-coded according to the local surface density. Stars excluded from the sample used for the detection of the RGB tip are plotted in greyscale, while those retained are plotted in Viridis scale.}
\label{fig:maps}
\end{figure*} 

In selecting the stars to be used for a clean detection of the RGB tip in the LMC and SMC H23 excluded from the sample the regions of the OGLE-III photometric maps where recent or ongoing star formation can inject into the color-magnitude diagram (CMD) young core He-burning stars that can contaminate the RGB, making the detection of the tip more uncertain (see Appendix~\ref{app:diff_hoyt} for further details and discussion).
Following this approach, we adopt a simple selection in angular distance from the galaxy center ({ see below}) and in interstellar extinction, taking advantage of the fact that the young stellar populations are concentrated in the central regions of both galaxies and that they are typically associated to higher concentrations of dust, hence high extinction.

For both galaxies, our starting sample is much more extended in area than the OGLE-III { footprint} and contains a much larger number of bright stars. For example, for LMC our original sample contains approximately two times the stars in the OGLE-III sample for $G<17.65$ \citep[the magnitude limit of GSPC][]{dr3_dpacp93}. Hence we were allowed to be more conservative than H23 in keeping only the less extincted { and less contaminated by young populations} outer regions of the galaxies.

In particular, we adopted the following selection criteria:

{
\begin{itemize}
    \item LMC: we retain only 202259 stars having E(B-V)$<0.2$ and with an angular distance larger than $3.8\degr$ 
    from (ra,dec)=$(81.28\degr,-69.00\degr)$
    \item SMC: we retain only 43152 stars having E(B-V)$<0.2$ and $R_{ang}>1.8\degr$ from (ra,dec)=$(15.00\degr,-72.70\degr)$
\end{itemize}

\noindent
In both cases the radial selection was not performed with respect to the centre of the galaxies \citep[as provided, e.g., by][]{edr3_luri}, but with respect to an off-set position optimising the removal of star forming regions from the sample with a simple circular cut. We do not attempt to correct for the 3d structure of the two galaxies, that is known to have non-negligible effects in both cases \citep[see, e.g.,][]{lmc_warp,edr3_luri,zivick21,okumar21,tatton21, cullinane2023,SMC_is_two}. Our approach is to average out these effects by means of large samples, with reasonably balanced distributions in azimuth. In Appendix~\ref{app:geo} we provide evidence supporting the validity of this choice.} 

The results of the selection are illustrated in Fig.~\ref{fig:maps}, where the finally selected samples are plotted in the Viridis density scale. For both galaxies ample fractions of the regions selected by H23 for his calibration are included in our samples. For the SMC, we decided to keep in the sample stars in the Magellanic Bridge \citep[see][for references]{edr3_luri} as we verified that the actual contamination of the RGB by young stars from this relatively small low-surface-density region is negligible.

Drawing from the GSPC provided for all the selected stars {\em standardised} magnitudes, fluxes and uncertainties on the fluxes in the following passbands: JKC UBVRI, SDSS ugriz, Hubble Space Telescope (HST) 
ACS-WFC\footnote{Advanced Camera for Surveys (ACS) Wide Fied Camera (WFC) \url{https://www.stsci.edu/hst/instrumentation/acs}} F606W and F814W. Then we used GaiaXPy\footnote{\url{https://gaia-dpci.github.io/GaiaXPy-website/}} \citep{dr3_xp_fda,dr3_dpacp93} to get, for the same stars, standardised magnitudes and  fluxes + uncertainties in PS1 grizy\footnote{The { GSPC} is supposed to contain also standardised PS1 y photometry but in fact, because of an error in writing the final catalogue, non-standardised PS1 y was included instead. See \url{https://www.cosmos.esa.int/web/gaia/dr3-known-issues\#SyntheticPhotSP1}}, and non-standardised magnitudes and  fluxes + uncertainties in JWST-NIRCAM F070W and F090W (VEGAMAG), NGRST R062 and Z087 (VEGAMAG), and in Euclid-VIS I$_E$
\citep[ABMAG;][]{cuillandre24}.

The uncertainties on magnitudes ($\epsilon_{mag}$) have been obtained from the uncertainties on fluxes ($\epsilon_{flux}$) as 

\begin{equation}
    \epsilon_{mag} = 1.086 \frac{\epsilon_{flux}}{flux} .
\end{equation}

\noindent
Each magnitude has been corrected for extinction using the reddening laws from Table~6 of \citep{schlafly2011} with $R_V=3.1$, when available. { In particular we adopted \footnote{ The coefficients reported in Table~6 of \citet{schlafly2011} are valid only for non-renormalised E(B-V) values read from the original \citet{sfd98} maps. We corrected them to have $R_V=3.1$ in Landolt's V, making them suitable for properly scaled E(B-V) values. In this scale $R_B=4.1$.}:

$$A_V=3.100E(B-V)$$
$$A_I=1.701E(B-V)$$ 
$$A_{F814W}=1.725E(B-V)$$
$$A_i=1.920E(B-V)$$
$$A_z=1.428E(B-V)$$
$$A_y=1.229E(B-V)$$
}

For passbands not included in that table, namely those in the JWST-NIRCAM, NGRST and Euclid systems, we adopted the reddening laws computed by the interface to the Padua stellar models\footnote{\url{http://stev.oapd.inaf.it/cgi-bin/cmd}} \citep{bressan2012} for a G2V star adopting the \cite{cardelli89} extinction curve for $R_V=3.1$, in particular:
\\

$$A_{F070W} = 2.309E(B-V),$$
$$A_{F090W} = 1.472E(B-V),$$
$$A_{R062} = 2.628E(B-V),$$ 
$$A_{Z087} = 1.572E(B-V),$$ 
$$A_{{\rm I}_E} = 2.285E(B-V).$$

\noindent
Given the restrictive reddening constraints adopted in the selection of our sample, uncertainties in the $A_{\lambda}/E(B-V)$ coefficients as well as neglecting the colour dependencies of the reddening laws should have negligible effects on the final correction.

\begin{figure*}[ht!]
\center{
\includegraphics[width=\columnwidth]{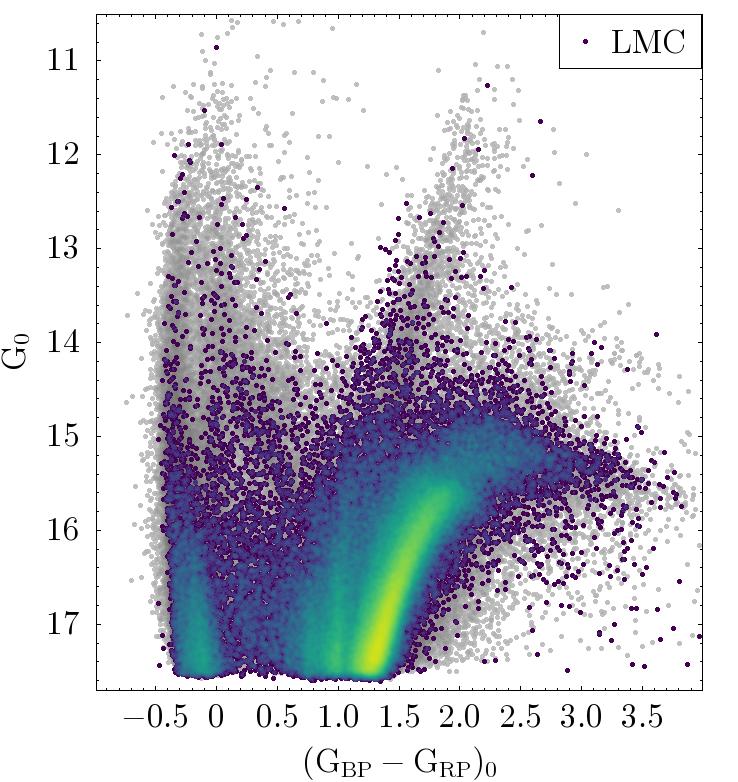}
\includegraphics[width=\columnwidth]{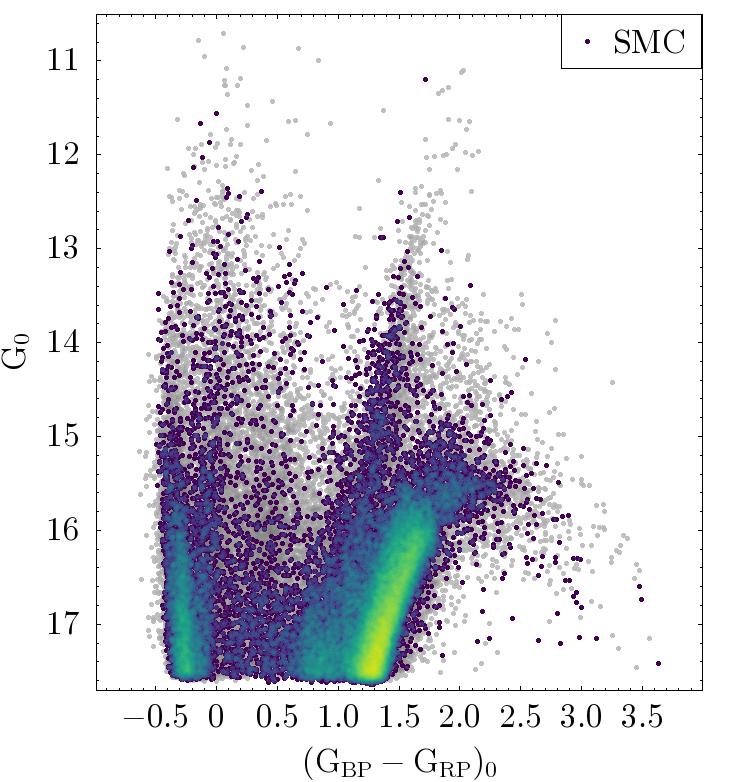}
}
\caption{CMD of the stars in our LMC sample (left panel) and SMC sample (right panel) in the Gaia DR3 photometric system. The meaning of symbols and colours is the same as in Fig.~\ref{fig:maps}.}
\label{fig:cmdsel}
\end{figure*} 

In Fig.~\ref{fig:cmdsel} we show the CMDs of the original (greyscale) and selected (viridis) samples for both galaxies. It is clear that large sets of genuine RGB stars can be very cleanly selected in the CMD, with very low degree of contamination from other stellar types near the RGB tip. A rich population of Asymptotic Giant Branch (AGB) stars is present above the tip, but the discontinuity in the RGB LF is clearly evident (around $G_0\simeq 15.5$ and  $G_0\simeq 15.8$ in the LMC and SMC, respectively), also in a passband showcasing a strong dependency of the RGB tip magnitude from colour, owing to the inclusion of a much bluer range of wavelength with respect to the red passbands that minimise these effects.
In the following, we select our sample of RGB stars for the detection of the tip with a simple selection in the CMD, by means of a pair of parallel lines enclosing the vast majority of RGB stars within a couple of magnitudes from the tip. We verified that the adopted selections in the various CMDs are consistent with one another, hence we are approximately using the same set of RGB stars in each passband / photometric system. For simplicity we decided to derive a single value of the RGB tip for each galaxy, associated to a mean colour, avoiding any colour splitting of the RGB of the LMC (see H23).

To transform extinction-corrected apparent magnitudes of the tip into absolute magnitudes we adopt the distance moduli from the eclipsing binaries measures, that is $(m-M)_0=18.477\pm 0.026$ for the LMC \citep{lmc19}, and $(m-M)_0=18.997\pm 0.032$ \citep{smc20}, where we summed in quadrature the statistic and systematic uncertainties quoted by these authors into a single uncertainty value.
In the following we will always use, as a reference, theoretical predictions for the magnitude of the RGB tip as a function of colour derived from PARSEC isochrones \citep{bressan2012}, obtained from the CMD~3.7 web interface\footnote{\url{http://stev.oapd.inaf.it/cgi-bin/cmd}}, as it provides stellar models in all the photometric systems considered here. 

\begin{table}[!htbp]
\caption{\label{tab:tipobs}. Extinction-corrected magnitude of the RGB tip of the MCs in  different passbands alongside the 1 and 3$\sigma$ confidence intervals (see Section~\ref{sec:model}).}
{
    \begin{tabular}{lccc}
gal &phot. sys.  &mag    &mag$_0^{TRGB}\pm 1\sigma(\pm 3\sigma)$\\  
\hline  

LMC &   JKC      &  I	 & 14.500  $^{+0.007(+0.020)}_{-0.006(-0.017)}$  \\ 
SMC &   JKC      &  I	 & 14.994  $^{+0.015(+0.051)}_{-0.015(-0.042)}$  \\ 
LMC & ACS-WFC    &F814W  & 14.493  $^{+0.006(+0.019)}_{-0.006(-0.018)}$  \\ 
SMC & ACS-WFC    &F814W  & 14.981  $^{+0.014(+0.047)}_{-0.014(-0.042)}$  \\ 
LMC &  SDSS	 &  i	 & 15.023  $^{+0.007(+0.021)}_{-0.006(-0.019)}$  \\ 
SMC &  SDSS	 &  i	 & 15.497  $^{+0.017(+0.059)}_{-0.015(-0.047)}$  \\ 
LMC &  SDSS	 &  z	 & 14.621  $^{+0.007(+0.024)}_{-0.007(-0.021)}$  \\ 
SMC &  SDSS	 &  z	 & 15.191  $^{+0.023(+0.072)}_{-0.022(-0.059)}$  \\ 
LMC &	PS1	 &  y	 & 14.530  $^{+0.008(+0.023)}_{-0.007(-0.021)}$  \\ 
SMC &	PS1	 &  y	 & 15.099  $^{+0.020(+0.068)}_{-0.017(-0.050)}$  \\ 
LMC &JWST-NIRCAM & F090W & 14.169  $^{+0.008(+0.025)}_{-0.007(-0.022)}$  \\ 
SMC &JWST-NIRCAM & F090W & 14.717  $^{+0.017(+0.054)}_{-0.016(-0.044)}$  \\ 
LMC &  NGRST	 & Z087  & 14.259  $^{+0.008(+0.025)}_{-0.007(-0.021)}$  \\ 
SMC &  NGRST	 & Z087  & 14.783  $^{+0.022(+0.074)}_{-0.021(-0.058)}$  \\ 
LMC &  Euclid-VIS    & I$_E$   & 15.230  $^{+0.005(+0.017)}_{-0.006(-0.017)}$  \\ 
SMC &  Euclid-VIS    & I$_E$   & 15.677  $^{+0.016(+0.053)}_{-0.014(-0.042)}$  \\ 
\hline
    \end{tabular}
}
\tablefoot{All the reported values are in units of magnitudes. The values of all the model parameters are reported in Table~\ref{tab:fitpar}. 
}
\end{table}

There are a few additional properties of the sample, inherent to the Gaia XPSP, that are worth reporting here. In \citet{dr3_dpacp93} the photometry in the JKC and SDSS systems has been standardised against large sets of reliable and well-tested standard stars, a large collection of Landolt's standard (\citealt{landolt92,landolt07b,landolt09,landolt13} and \citealt{landuomoto}), later  critically assessed by \citet{pancino22} and the most-recent Stripe~82 set by \citet{thanjavur2021}, respectively. Hence BVRI and griz XPSP magnitudes\footnote{The accuracy and precision of standardised XPSP magnitudes in the U$_{JKC}$ and u$_{SDSS}$ bands are significantly less good than for these passbands. See \citet{dr3_dpacp93} for a thorough discussion.} for these systems should be considered as the most reliably and robustly tested to reproduce the original systems with high accuracy. Magnitudes in the PS1 system (grizy) have also been standardised against a large sample of high-quality PS1 photometry but not from a ``standard set'', just a suitable patch of the survey catalogue \citep{Magnier2020}. Magnitudes in the ACS-WFC system have been standardised against a relatively small dataset of globular cluster stars \citep{nardiello18}, for reasons discussed in \citet{dr3_dpacp93}. XPSP photometry reproduces magnitudes in these systems with accuracy comparable to the two cases above but the standardisation is based on (slightly for PS1, significantly for ACS-WFC) less robust external datasets.  

XPSP for the JWST-NIRCAM, NGRST and Euclid-VIS systems is not standardised, as there were no external dataset to be used for this purpose. In this case the accuracy can be tentatively inferred from that measured by \citet{dr3_dpacp93} in standardised systems before standardization in passbands covering similar ranges of wavelength as those considered here for RGB tip calibration\footnote{The accuracy of XPSP as well as of any synthetic photometry depends also on the accuracy to which the transmission curves provided for the various passbands represent the real transmission curves. The standardisation process adopted by \citet{dr3_dpacp93} is intended to correct XPSP not only for systematics affecting externally calibrated XP spectra but also for inaccuracies of the adopted transmission curves.}. This is typically $\simeq 0.01$~mag but in some cases reaches $0.03-0.04$~mag: this is an intrinsic unavoidable uncertainty affecting the RGB tip calibrations derived below for magnitudes from non-standardised XPSP. { In Sect.~\ref{sec:calALL} we illustrate how the uncertainty in the photometric zero points can be taken into account in our final calibrations of the RGB tip.}
Non standardised Gaia XPSP is also affected by a trend with magnitude, the so called hockey stick effect, but its amplitude is negligible in the range of magnitudes where the MC tips lie, being $<0.01$ for $G\le 16.0$.

Finally, \citet{dr3_dpacp93} showed that the photometric uncertainty associated to each individual XPSP flux/magnitude is typically underestimated by a factor ranging from $\simeq 1$ up to $\simeq 5$, depending on the star magnitude. This effect has been corrected with an empirical formula provided in \citet{dr3_dpacp93} and included in GaiaXPy, for all the systems that were available in the GaiaXPy repository at that epoch. The individual photometric errors in our samples are corrected in this way for all the magnitudes except for Euclid-VIS I$_E$\footnote{The Euclid system was among those included in the original GaiaXPy list but it was erroneously computed as a VEGAMAG system.}. 

\section{The RGB tip of the LMC and SMC from XPSP: validation and methods}
\label{sec:measure}

\begin{figure*}[ht!]
\center{
\includegraphics[width=\columnwidth]{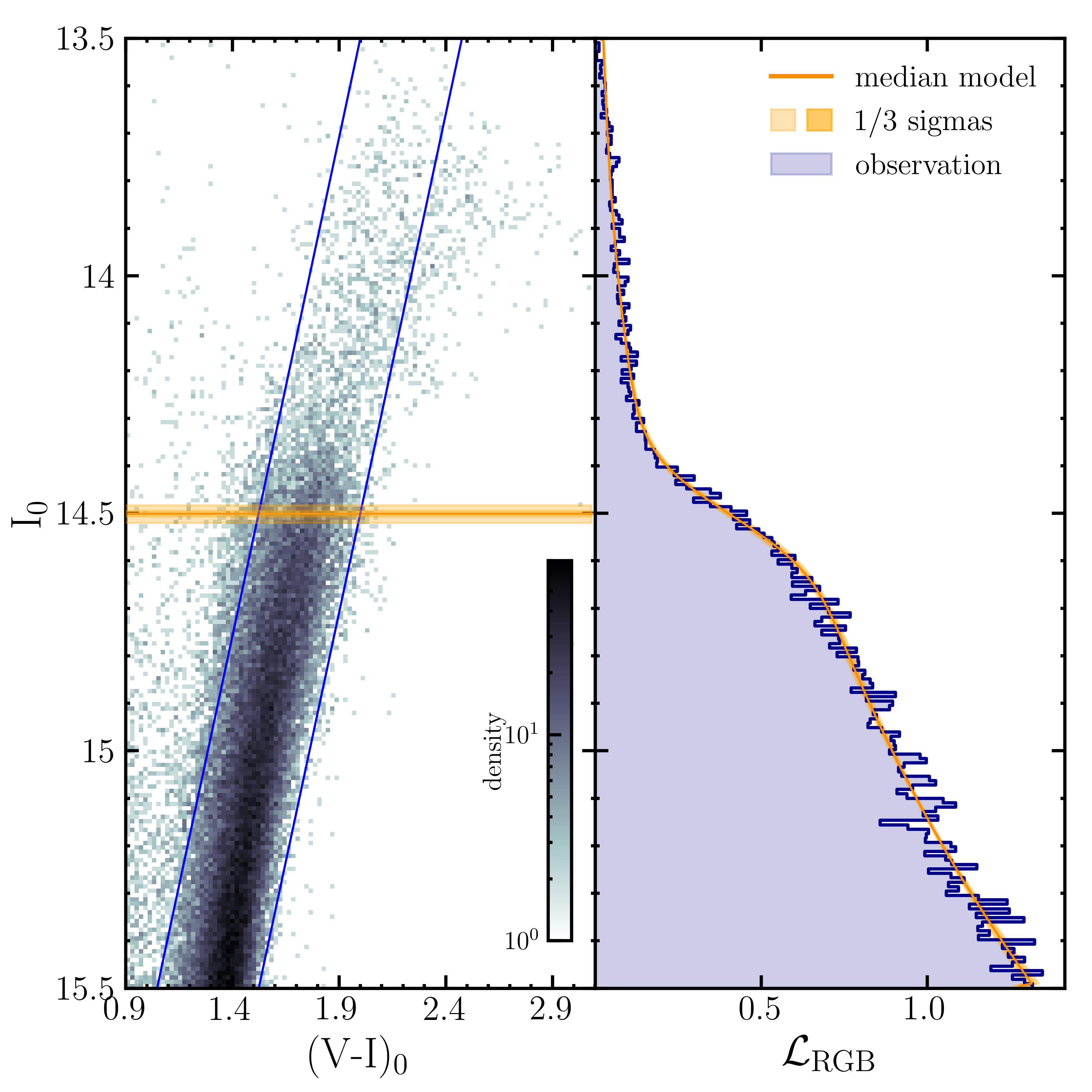}
\includegraphics[width=\columnwidth]{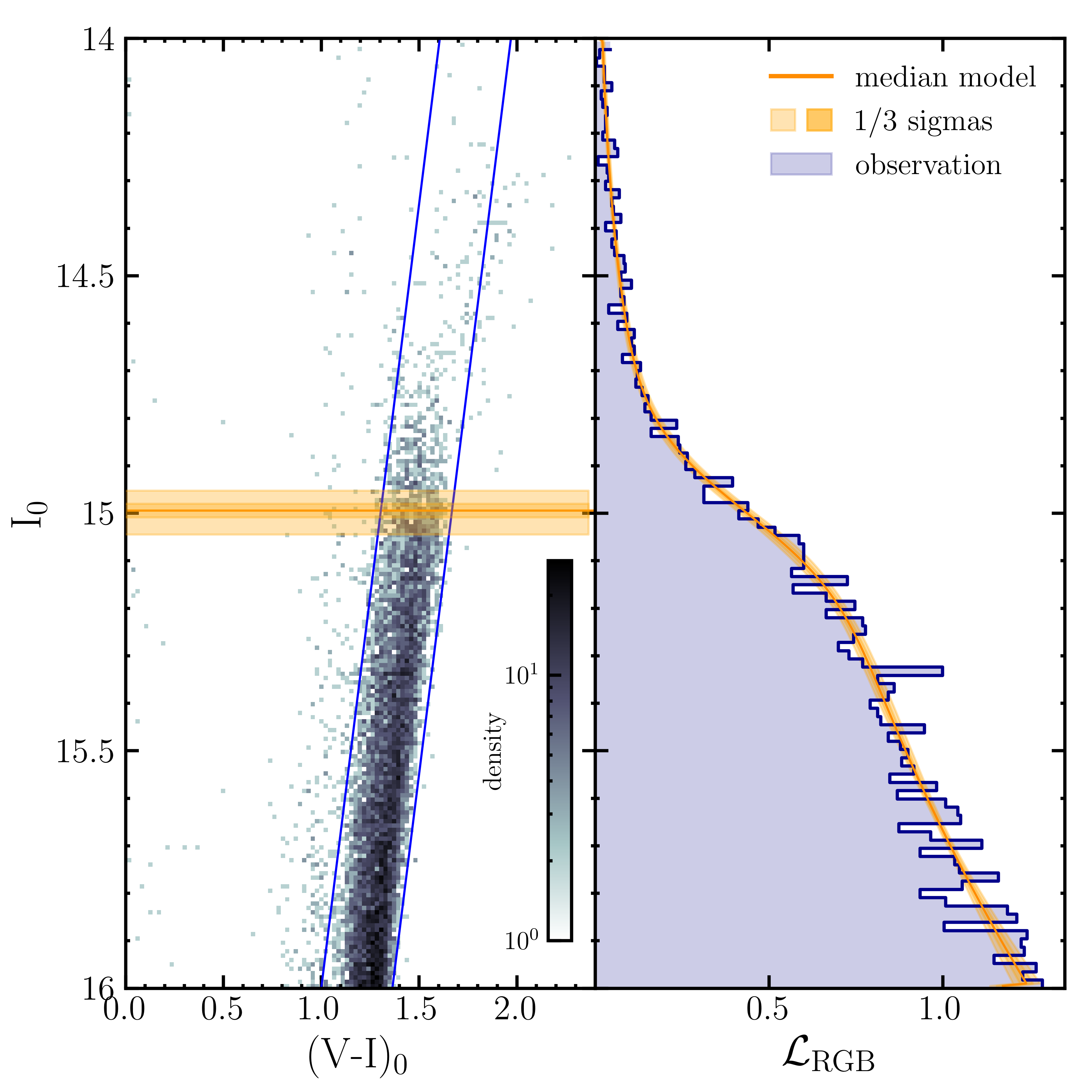}
}
\caption{Measuring the RGB tip in JKC I band in the LMC (left pair of panels) and in the SMC (right pair of panels). The left panels of each pair display the CMD of the considered sample with the selection of the RGB + AGB sub-sample that is later used to construct the LF (stars enclosed within the blue parallel lines). The orange horizontal line marks the position of the RGB tip and the shaded bands show the $1\sigma$ and $3\sigma$ confidence intervals. The right panels of each pair show the LF of the selected RGB + AGB sub-sample (light blue histogram) and the best fit model by which we estimate $I^{TRGB}$.}
\label{fig:vitip}
\end{figure*}

In Fig.~\ref{fig:vitip} the measure of the RGB tip in the LMC and SMC in the JKC I band is displayed. The measure is performed fitting to the RGB LF a model similar to those of \citet{mendez02, makarov06,conn11,conn12,conn16}. The details of our version of this method are described in Sect.~\ref{sec:method}.

The RGB tip is very cleanly detected in both galaxies at $I_0^{TRGB}=14.500_{-0.006(-0.017)}^{+0.007(+0.020)}$ for the LMC and at 
$I_0^{TRGB}=14.994_{-0.015(-0.042)}^{+0.5(+0.051)}$ for the SMC, where the reported uncertainties are proxies for the $1(3)\sigma$ confidence intervals (in particular, they are the 16th(0.15th) and 84th(99.85th) percentiles of the posteriori distribution of $I_0^{TRGB}$; see Sect.~\ref{sec:method}). The colour of the RGB tip corresponding to the aforementioned measurements is estimated as the median colour of the RGB stars with a magnitude within the $1\sigma$ confidence interval of the tip magnitude. The associated uncertainty is calculated as the 16th and 84th percentiles of the colour distribution of stars with that magnitude. 
In all the cases presented in the following analysis, this is the method employed to evaluate the colour of the RGB tip.

As a consistency check, H23 computed the difference between the measured RGB tip of the LMC and the SMC, obtaining $\Delta \mu=0.488 \pm 0.024$ ($\Delta \mu=0.494$ applying the colour correction by \citealt{jang17}, accounting for the colour difference between the RGB tip of the two galaxies), in excellent agreement with the difference between the distance moduli measured with eclipsing binaries $\Delta \mu=0.500 \pm 0.017$ \citep{lmc19,smc20}. From our measures we obtain
$\Delta \mu=0.494 \pm 0.016$ ($\Delta \mu=0.506$ applying the colour correction), in excellent agreement with both the H23 result and the reference value from eclipsing binaries and more closely matching the latter value than what obtained by H23.
On the other hand our measures of the RGB tip are fainter that those obtained by H23 by $0.061\pm 0.009$~mag for the LMC (formally a full $6.8\sigma$ difference) and by $0.067\pm 0.027$~mag for the SMC ($2.5\sigma$). In Appendix~\ref{app:diff_hoyt} we explore in detail the possible reasons for this discrepancy. 

\begin{figure}[ht!]
\includegraphics[width=1\hsize]{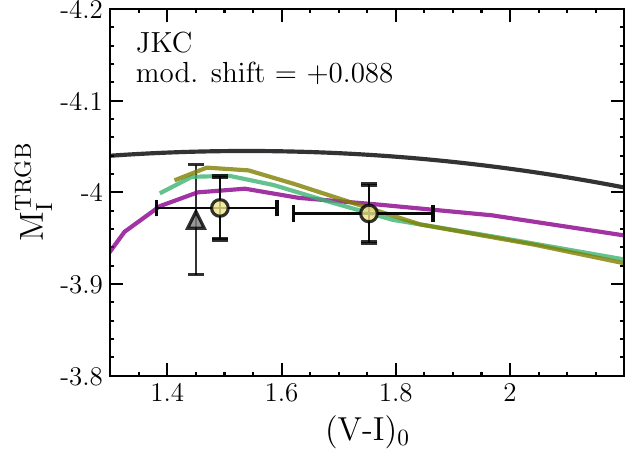}
\caption{I-band absolute magnitude of the RGB tip as a function of V-I colour. The calibration of our measures in the SMC and LMC are plotted as yellow filled circles with vertical error-bar at $1\sigma$ and $3\sigma$ confidence intervals and horizontal error-bars at $1\sigma$ confidence interval. In all the cases the point corresponding to the SMC is to the blue of the point corresponding to the LMC. The grey triangle is the calibration obtained by \citet{soltis21} in $\omega$~Cen. The black curve is the calibration provided by H23 in form of a second order polynomial fit. The coloured curves are model predictions obtained from the set of Padua isochrones by \citet{bressan2012}, corresponding to ages of 4~Gyr (purple line), 10~Gyr (light blue), and 13~Gyr (green line), all vertically shifted by the amount reported in the plot (mod. shift). The shift is applied to make the 10~Gyr model, taken as reference, to match the point corresponding to the LMC.}
\label{fig:mycalVI}
\end{figure}

H23 showed that many recent calibrations of the RGB tip in I band have zero-point (ZP)
in the range between $M_I^{TRGB}\simeq -4.00$ and $M_I^{TRGB}\simeq -4.05$ { \citep[see, e.g.,][for a recent example]{dixon23}}, with a few cases in which fainter ZP are obtained \citep[$M_I^{TRGB}\simeq -3.97$;][]{yuan19, soltis21, li23}. H23 argued that the latter calibrations are based on faint measures of the tip of the LMC due to improper correction of the extinction and (above all) selection of the fields, with respect to his analysis (in Appendix~\ref{app:diff_hoyt} we address this specific point in our case). Fig.~\ref{fig:mycalVI} shows very clearly that, due to the differences in $I_0^{TRGB}$ mentioned above, the ZP of our calibration of the RGB tip ($M_I^{TRGB}=-3.98\pm 0.03$ for the LMC) is indeed much more similar to those obtained by \citet{yuan19}, \citet{soltis21}, and \citet{li23} than to the ``bright'' set of ZPs, including H23 and \citet{mbtip01}. { However, in Appendix~\ref{app:diff_hoyt} we  show that, once the differences in photometric zero points are considered, the actual disagreement with \citet{hoyt23} is mitigated and the match with \citet[][Y19 hereafter]{yuan19} is, in fact, less close than how it appears at face value. It may also be worth noting that the calibration by \citet{li23} is partially based on Gaia DR3 GSPC photometry, hence is not completely independent with respect to our measure.} 

In Fig.~\ref{fig:mycalVI}, our measures are represented as yellow filled circles, the one by \citet{soltis21} for $\omega$~Cen by a grey filled triangle, and the black curve is the quadratic polynomial relation by \citet{jang17} with the ZP adjusted to match the measures by H23, that is a good synthetic representation of the H23 results over a large colour range.
It is interesting to note that, in spite of the obvious difference in the ZP, the branch of the curve in the range of colours where our points are located is very flat, in excellent agreement with the virtually null difference that we obtain between $M_I^{TRGB}$ of the LMC and of the SMC.

In Fig.~\ref{fig:mycalVI}, as well as in all the following analogue diagrams, we plot the theoretical predictions for the absolute magnitude of the tip as a function of colour for solar-scaled models of age 4~Gyr (purple line), 10~Gyr (light blue line), and 13~Gyr (green line). 
The 13~Gyr and 4~Gyr models are intended to provide a quantitative view of the amplitude of the dependency on age of the standard candle, in this case $\la 0.02$~mag at any colour, while the 10~Gyr model is taken as a "mean" reference to give an idea of the predicted trend with colour outside the range covered by our measured points, mainly toward redder colours / higher metallicities. All the model predictions have been shifted by the amount needed to reach a match between the observed $M_I^{TRGB}$ of the LMC and the prediction of the 10~Gyr model at that colour. In the present case the shift, labelled as "mod. shift" in this kind of plots, is +0.088~mag, indicating that the predictions of PARSEC models are in better agreement with the H23 ZP, and in general to the bright set of RGB tip calibrations, that with our calibration or of those by \citet{soltis21} and \citet{li23}. In fact, they are 0.01~mag to 0.06~mag brighter also than the quadratic polynomial based on the H23 calibration, depending on the colour and age of the models.
It is also worth noting that the dependency on colour of the age $\la10$~Gyr models is quite different from the polynomial representing the H23 calibration, especially in the blue / metal-poor regime around the colour of the SMC. Stellar models have their own limitations that should be always considered in this kind of comparisons.

In summary, we use a large and independent sample of LMC and SMC stars with accurate Gaia standardised VI XPSP\footnote{That is reproducing the JKC system photometry as defined by Landolt's standard 
stars (see \citealt{landuomoto},
\citealt{landolt92,landolt07b,landolt09,landolt13}  
and also \citealt{pancino22}) with a few millimag accuracy \citep{dr3_dpacp93}}, 
state-of-the-art extinction corrections, and the distance moduli from 
\citet{lmc19} and \citet{smc20}, to obtain a remarkably precise measure of $M_I^{TRGB}$ for the LMC and SMC. The comparison of our results with the most recent and most accurate calibrations presented above shows that our result is $\simeq 0.05$~mag fainter that most of them, but in agreement with others. This is what the user of the calibrations in the other photometric systems presented in Sect.~\ref{sec:calALL} must keep in mind as it put our measures in the context of the relevant literature in the system for which this literature exist. The comparison presented in this section should be considered as a validation and an assessment of the accuracy of the results that can be obtained with the same samples and the same methods in the other photometric systems considered here.

\subsection{Calibration of the RGB tip in different systems}
\label{sec:calALL}

The detection of the TRGB in the LMC and SMC samples for the various passbands considered here is presented in Appendix~\ref{app:dete}, using diagrams fully analogue to those shown in Fig.~\ref{fig:vitip} for the JKC I band. The resulting measures of the extinction-corrected apparent magnitude of the tip in the various passbands for the LMC and the SMC, alongside the associated $1\sigma$ and $3 \sigma$ confidence intervals (as defined in Sect.~\ref{sec:measure}), are listed in Table~\ref{tab:tipobs}. 
 
Here we show the calibration of the absolute magnitude of the RGB tip as a function of colour adopting the same arrangement as in Fig.~\ref{fig:mycalVI} for the following magnitudes (passbands): ACS-WFC F814W, PS1 y, SDSS i and z, JWST-NIRCAM F090W, NGRST Z087, and Euclid-VIS I$_E$. The choice of the colour was in some case natural for the general use (F606W-F814W for the ACS-WFC system) or for the availability of only two passbands enclosed in the XP spectral range (JWST-NIRCAM and NGSRT systems). For the SDSS system we selected the g-i colour for its high sensitivity to temperature while to display the colour dependence of the tip in Euclid-VIS I$_E$ we were forced to recur to JKC V-I colour, as I$_E$ is the only Euclid passband enclosed within the XP spectral range. 
 
 All the plots have the same scale in the y-axis, for a direct and easy comparison of the amplitude of the dependencies on colour (metallicity) and age, as predicted by the adopted set of stellar models. The only exception is the case of Euclid-VIS I$_E$, where a larger scale was adopted to accommodate a significantly stronger dependency on colour, that is not surprising given the much bluer wavelength range of this passband with respect to all the other ones.
On the other hand the colour range has been chosen to enclose the range spanned by the 10~Gyr model for metallicity $-2.0\le$[M/H]$\le -0.4$. The colour and the absolute magnitude of the RGB for the various passbands and for the LMC and SMC are reported in Table~\ref{tab:tipcal}.

\begin{figure}[ht!]
\includegraphics[width=1\hsize]{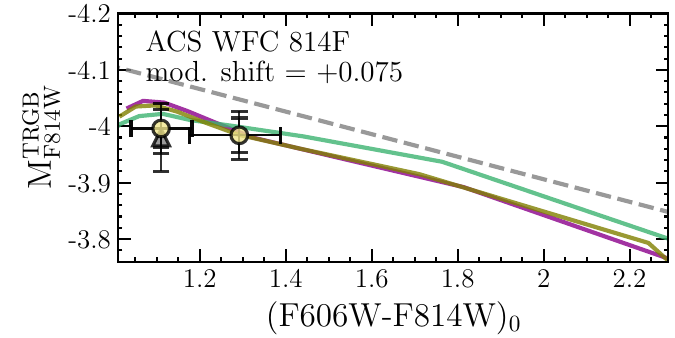}\\
\includegraphics[width=1\hsize]{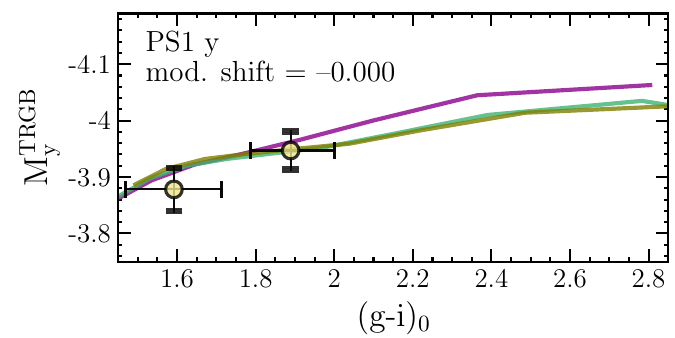}
\caption{Calibration of the RGB tip as a function of colour. Upper panel:
absolute magnitude of the RGB tip in ACS-WFC F814W band as a function of F606W-F814W colour. The dashed line is the linear calibration by \citet{rizzi07}. Lower panel: absolute magnitude of the RGB tip in PS1 y band as a functon of PS1 g-i colour. The meaning of the symbols and the arrangement are the same as in Fig.~\ref{fig:mycalVI}.}
\label{fig:mycal1}
\end{figure} 

In the upper panel of Fig.~\ref{fig:mycal1} we show $M_{F814W}^{TRGB}$ as a function of $(F606W-F814W)_0$ in the ACS-WFC system. F814W is very similar to JKC I and the results are indeed very similar to that shown in Fig.~\ref{fig:mycalVI}, above. Here we also show the comparison with the linear calibrating relation by \citet{rizzi07} that has a ``bright'' ZP, similar to H23.  In the lower panel of Fig.~\ref{fig:mycal1}, concerning the PS1 y band, it can be appreciated that the $M_{y}^{TRGB}$ difference between the SMC and the LMC broadly follows the trend predicted by the models\footnote{It is interesting to note that this is the case in which the difference between the tip of the SMC and the LMC is most significant, albeit just $\simeq 1.5\sigma$, see below.}. The colour dependency is slightly lower than for the JKC I band, in the considered colour range, while the dependency on age has a similar amplitude.

\begin{figure}[ht!]
\includegraphics[width=1\hsize]{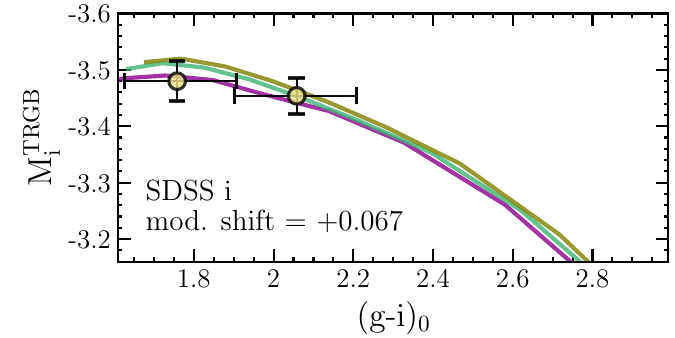} \\
\includegraphics[width=1\hsize]{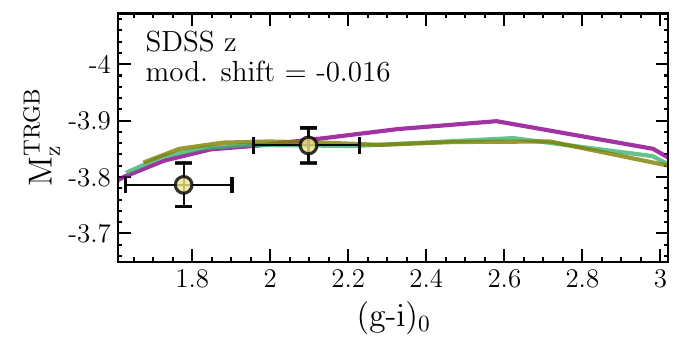}
\caption{Calibration of the RGB tip as a function of colour. Upper panel:
absolute magnitude of the RGB tip in SDSS i band as a function of SDSS g-i colour. Lower panel: absolute magnitude of the RGB tip in SDSS z band as a function of SDSS g-i colour. The meaning of the symbols and the arrangement are the same as in Fig.~\ref{fig:mycalVI}.}
\label{fig:mycal2}
\end{figure} 

Fig.~\ref{fig:mycal2} shows that SDSS i has a stronger dependency on colour than JKC I while it displays the smallest amplitude of age dependency of all the passbands considered here. However, in contrast with all the other cases, the sensitivity to age is maximal in the metal-poor (blue) regime.  In the lower panel of the same figure it can be appreciated the very low sensitivity of $M_{z}^{TRGB}$ to colour, according to the PARSEC model, over the entire colour range considered. 

\begin{figure}[ht!]
\includegraphics[width=1\hsize]{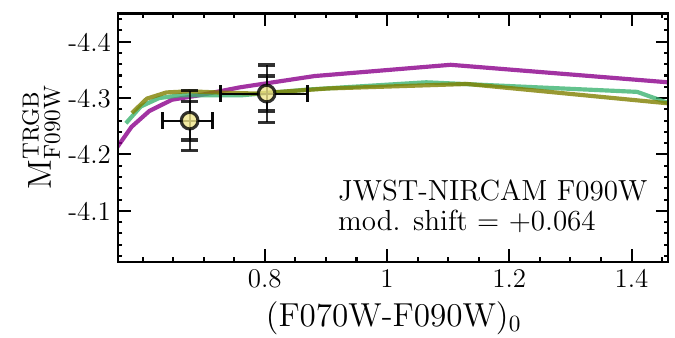} \\
\includegraphics[width=1\hsize]{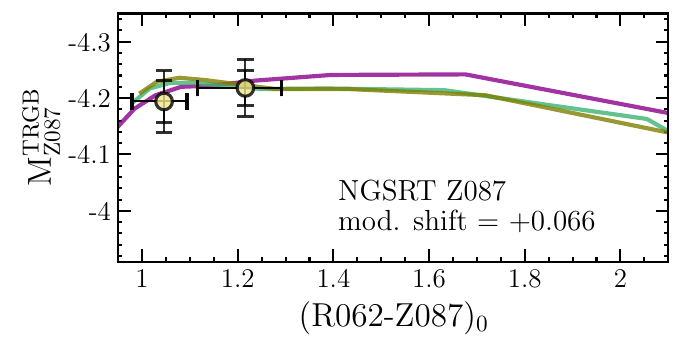}
\caption{Calibration of the RGB tip as a function of colour. Upper panel:
absolute magnitude of the RGB tip in JWST-NIRCAM F090W band as a function of F070W-F090W colour. Lower panel: absolute magnitude of the RGB tip in NGRST Z087 band as a function of R062-Z087 colour. The meaning of the symbols and the arrangement are the same as in Fig.~\ref{fig:mycalVI}.}
\label{fig:mycal3}
\end{figure}

JWST-NIRCAM $M_{F090W}^{TRGB}$ and NGRST $M_{F087}^{TRGB}$ appear similarly well-behaved, as it can be appreciated from Fig.~\ref{fig:mycal3}. It is interesting to note that, in a very recent paper, \citet{anand24} provided a calibration of the RGB tip in the JWST-NIRCAM F090W band from a measure in a galaxy at $D\simeq 7.6$~Mpc for which a geometric distance is available, NGC~4258, whose mean colour of the tip is quite similar to that of the LMC\footnote{In the F814W-F606W colour, see \citet[][]{li23}.}. They find $M_{F090W}^{TRGB} = -4.362\pm 0.033{\rm (stat)} \pm 0.045{\rm (sys)}$, { compatible} with our value for the LMC $M_{F090W}^{TRGB} = -4.308\pm 0.051$ ({ where the reported errorbar includes also the contribution of the uncertainty in the photometric zero point;} see Table~\ref{tab:tipcal}), thus providing a valuable cross-validation of the two (fully independent) calibrations in this new photometric system.

\begin{figure}[ht!]
\includegraphics[width=1\hsize]{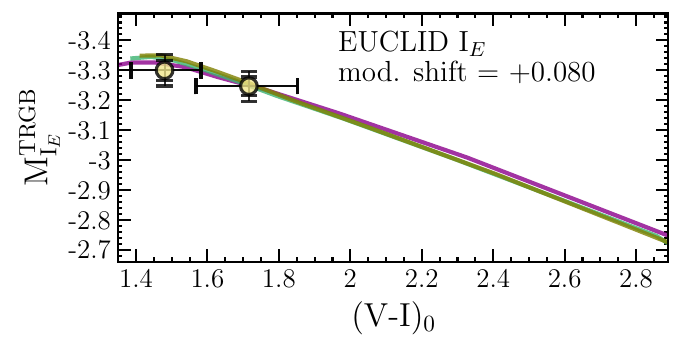}
\caption{Calibration of the RGB tip as a function of colour. 
Absolute magnitude of the RGB tip in Euclid-VIS I$_E$ band as a function of JKC V-I colour. The meaning of the symbols and the arrangement are the same as in Fig.~\ref{fig:mycalVI} except for the scale of the y axis, that here is wider to accommodate for the larger amplitude of the $M_{{\rm I}_E}^{TRGB}$ trend with colour, with respect to the other considered passbands.}
\label{fig:mycal4}
\end{figure}

As expected, the dependency of  $M_{{\rm I}_E}^{TRGB}$ on colour is large when the entire range spanned in Fig.~\ref{fig:mycal4} is considered. However, for colours bluer than the LMC tip the amplitude of the predicted trend is $\la 0.1$~mag and the difference in $M_{{\rm I}_E}^{TRGB}$ between the LMC and the SMC is just $0.05\pm 0.04$, suggesting that it may be a good standard candle in this blue colour range.

In general the difference in $M_{mag}^{TRGB}$ between SMC and LMC is never found to be statistically significant. The observed differences (SMC-LMC) are $-0.13\sigma$, $-0.27\sigma$, $-0.57\sigma$, $1.43\sigma$, $1.47\sigma$, $1.04\sigma$, $0.50\sigma$, and 
, $-1.17\sigma$, for JKC I, ACS-WFC F814W, SDSS i, SDSS z, PS1 y, JWST-NIRCAM F090W, NGRST Z087, and Euclid-VIS I$_E$, respectively. In this sense, those interested to use our calibrations can take, as a reference value, $M_{mag}^{TRGB}$ of the galaxy, between SMC and LMC, having the colour more similar to their target stellar system, or the weighted mean of the two values, if their target has colour within the range spanned by the SMC and LMC tips. To extrapolate to significantly redder colours one can recur to models, re-adjusting the ZP to match our measures as we did above, or in cases of very weak metallicity dependency, like e.g., $M_{z}^{TRGB}$, $M_{F090W}^{TRGB}$ or $M_{Z087}^{TRGB}$, taking our measures as a reference in the entire range of colours spanned by the plots presented in this section. According to the considered models, this should lead to very small systematics, with amplitude of the same order { as} the age dependency ($\la 0.03$ mag).  

{ In Appendix~\ref{app:diff_hoyt} we provide some examples of the impact that the uncertainty in the photometric zero point may have in the total error budget\footnote{ Please note that, in the determination of the distance of a given stellar system, this source of uncertainty plays a twofold role, as it may independently affect the calibration of the adopted standard candle and the observations of the candle in the target system.}. Given the results presented in Appendix~\ref{app:diff_hoyt} and, in particular, the arguments presented at the last paragraphs of Sect.~\ref{sec:samples}, we decided to provide, in Table~\ref{tab:tipcal}, the total error including also the contribution of this source of uncertainty.
Following the discussion in Sect.~\ref{sec:samples} \citep[see][]{dr3_dpacp93} we conservatively adopted a contribution from this factor of 0.01~mag for the calibrations in the JKC and SDSS systems, of 0.02~mag for the PS1 system, 0.03~mag for the HST ACS-WFC system, and 0.04~mag for the remaining, non standardised, systems. It is worth noting that in all the considered cases the total uncertainty on $M_{mag}^{TRGB}$ remains $\la 0.06$~mag.
We recommend to adopt this global error when using our calibrations, except for cases in which there is a good reason not to do so, like, for instance when the distance to a stellar system is derived from Gaia XP synthetic photometry, as we do here when comparing $M^{TRGB}_{mag}$ of the LMC and SMC. For this kind of applications, in Tab.~\ref{tab:tipcal} we provide also the version of the uncertainty on $M_{mag}^{TRGB}$ without the contribution of the uncertainty on the photometric zero point.}

The relevant point here is that in Table~\ref{tab:tipcal} we provide robust estimates of the absolute magnitude of the RGB tip for two fundamental pillars of the cosmological distance ladder in several suitable and widely used passbands for which such calibration was not available before. Moreover, we used one set of stellar model to provide an idea of the behaviour of the newly calibrated standard candles in colour ranges not covered by our measures. These tools should be sufficient to get distance estimates to stellar systems with well populated RGBs with accuracy better { than} 5-10\% in all the considered photometric systems and in a wide variety of cases.

\begin{table*}[!htbp]
\centering
\caption{\label{tab:tipcal}. Colour and absolute magnitude of the RGB tip of the MCs in 
different passbands alongside $1\sigma$ confidence intervals.}
{
    \begin{tabular}{lccccc}
\hline
gal &photometric system  &mag    &colour & ${\rm col}_{0}$      & $M_{mag}^{TRGB}$\\  
\hline  
LMC &   JKC      &  I	 &V-I            & 1.753$_{-0.132}^{+0.112}$ & -3.977 $\pm$0.032 (0.031) \\
SMC &   JKC      &  I	 &V-I            & 1.492$_{-0.111}^{+0.100}$ & -3.983 $\pm$0.035 (0.033) \\
LMC & ACS-WFC    &F814W  &F606W-F814W    & 1.292$_{-0.116}^{+0.095}$ & -3.984 $\pm$0.043 (0.030) \\
SMC & ACS-WFC    &F814W  &F606W-F814W    & 1.110$_{-0.070}^{+0.072}$ & -3.996 $\pm$0.045 (0.033) \\
LMC &  SDSS	 &  i	 &(g-i)$_{SDSS}$ & 2.058$_{-0.155}^{+0.149}$ & -3.454 $\pm$0.032 (0.031) \\
SMC &  SDSS	 &  i	 &(g-i)$_{SDSS}$ & 1.758$_{-0.133}^{+0.150}$ & -3.480 $\pm$0.036 (0.034) \\
LMC &  SDSS	 &  z	 &(g-i)$_{SDSS}$ & 2.098$_{-0.141}^{+0.130}$ & -3.856 $\pm$0.032 (0.031) \\
SMC &  SDSS	 &  z	 &(g-i)$_{SDSS}$ & 1.778$_{-0.150}^{+0.124}$ & -3.786 $\pm$0.039 (0.038) \\
LMC &	PS1	 &  y	 &(g-i)$_{PS1}$  & 1.889$_{-0.102}^{+0.112}$ & -3.947 $\pm$0.037 (0.031) \\
SMC &	PS1	 &  y	 &(g-i)$_{PS1}$  & 1.592$_{-0.123}^{+0.122}$ & -3.877 $\pm$0.041 (0.036) \\
LMC &JWST-NIRCAM & F090W & F070W-F090W   & 0.803$_{-0.075}^{+0.067}$ & -4.308 $\pm$0.051 (0.031) \\
SMC &JWST-NIRCAM & F090W & F070W-F090W   & 0.677$_{-0.044}^{+0.038}$ & -4.260 $\pm$0.053 (0.034) \\
LMC &  NGRST	 & Z087  & R062-Z087     & 1.216$_{-0.100}^{+0.075}$ & -4.218 $\pm$0.051 (0.031) \\
SMC &  NGRST	 & Z087  & R062-Z087     & 1.046$_{-0.067}^{+0.048}$ & -4.194 $\pm$0.055 (0.037) \\
LMC &  Euclid-VIS    & I$_E$   &  V-I          & 1.716$_{-0.148}^{+0.136}$ & -3.247 $\pm$0.050 (0.030) \\
SMC &  Euclid-VIS    & I$_E$   &  V-I          & 1.480$_{-0.094}^{+0.102}$ & -3.300 $\pm$0.052 (0.034) \\
\hline
    \end{tabular}
}
\tablefoot{All the reported values are in units of magnitudes. col$_0$ is the reddening-corrected median colour of 
the RGB tip. The colour for the RGB tip in Euclid-VIS I$_E$ is JKC V-I; 
The uncertainties on $M_{mag}^{TRGB}$ have been computed 
summing in quadrature the global uncertainty on the distance modulus, the largest between the $-1\sigma$ and $+1\sigma$ 
uncertainty on mag$_0^{TRGB}$ from Tab.~\ref{tab:tipobs}, and the adopted uncertainty in the photometric zero point. 
The value within parentheses is the uncertainty when the latter factor is neglected.}
\end{table*}

\section{The adopted tip detection algorithm}
\label{sec:algo}

In this Section, we outline the model used to describe the RGB LF (Section~\ref{sec:model}) and the methodology employed for fitting it to a given dataset, with the aim to determine the RGB tip (Section~\ref{sec:method}).
There are two kind of methods by which the measure of the RGB tip has been performed in the literature, the use of an edge detector filter \citep[typically the Sobel's filter,][H23]{lee93,madore95,sakai96} or the fitting of a simple model to the brightest portion of the RGB LF plus an additional component accounting for AGB stars brighter that the tip and back/foreground contaminating populations \citep{mendez02,makarov06,conn11,conn12,conn16}. The first method, in principle, is simple and non-parametric but the response of the edge-detector filter can be very noisy, especially in cases of sparsely populated LFs, giving rise to some ambiguity in the identification of the peak of the filter response that is actually associated to the RGB tip. Moreover, the error associated with the measure is not well-defined (see H23 for an empirical approach based on re-sampling). The second method can significantly mitigate the effects of shot noise and provides a proper quantitative description of uncertainties and covariances \citep[see,e.g.][]{conn12}. A potential drawback is that the measure of mag$_0^{TRGB}$ can be somehow influenced by the use of a model that is not flexible enough to fit simultaneously all the components of the RGB LF (e.g., its bright/faint end, the sharpness of the transition around the tip magnitude, etc.). Here, we opted for a model-fitting technique, above all for the excellent control of the uncertainties it provides. 

\subsection{The model}
\label{sec:model}

In the adopted approach, the RGB LF, $\LLRGB$, in a given photometric band, is described by
\begin{equation}\label{for:model}
\LLRGB(m) = \left\{
\begin{array}{ll}
    \frac{\LLs(m)}{\int_{\mdown}^{\mup}\LLs(m)\dd m}  & \mdown\le m\le\mup \\
    0  & {\rm otherwise}, \\
\end{array}\right.
\end{equation}
where $[\mdown,\mup]$ marks the bandwidth of the considered portion of the upper RGB. In equation~(\ref{for:model}), $\LLRGB$ is normalized to unity and $\LLs$ is computed applying a Gaussian smoothing to
\begin{equation}\label{for:lfdisc}
\LL(m) = \left\{
\begin{array}{lr}
   10^{a(m-\mtip)+c}  & m\ge\mtip \\
   10^{b(m-\mtip)} & m<\mtip.
\end{array}\right.
\end{equation}
where the $m\ge\mtip$ branch is intended to model the AGB  population. This model was previously used, for instance, by \cite{makarov06,conn12,conn16} to describe the RGB LF, in the context of the determination of the RGB tip. Therefore,
\begin{equation}\label{for:smooth}
    \LLs(m) = (\LL \circledast G)(m) = \int_{-\infty}^{+\infty} \LL(x)G(x-m)\dd x,
\end{equation}
with $\circledast$ the convolution operator and $G$ a Gaussian of null mean and $\sigtip$ dispersion. It can be shown that the analytical expression for equation~(\ref{for:smooth}) is
\begin{equation}\label{for:final}
\begin{split}
    \LLs(m) = & 10^{a(m-\mtip)+c+\frac{a^2\sigtip^2\ln10}{2}} \times \\
    & \biggl[\erf\bigg(\frac{m-\mtip}{\sqrt{2}\sigtip} + \frac{a\sigtip\ln10}{\sqrt{2}}\biggr) + 1\biggr] + \\
    & 10^{b(m-\mtip)+\frac{b^2\sigtip^2\ln10}{2}} \times \\ &
    \biggl[1-\erf\bigg(\frac{m-\mtip}{\sqrt{2}\sigtip} + \frac{b\sigtip\ln10}{\sqrt{2}}\biggr)\biggr],
\end{split}
\end{equation}
where $\erf(x)$ is the error function and $\mtip$ the magnitude of the RGB tip, that is the main observable quantity we want to measure. 

For $m\gg\mtip$, $\log\LL(m)$ is a power-law of slope $a$, while for $m\ll\mtip$, $\log\LL(m)$ is a power-law of slope $b$. The LF~(\ref{for:lfdisc}) is discontinuous at $m=\mtip$ ($\mdown\le\mtip\le\mup$), with $c$ describing the amplitude of the discontinuity. The use of equation~(\ref{for:final}) allows for a continuous modeling of the luminosity transition around the tip magnitude, but keeping the same asymptotic behaviors of model~(\ref{for:lfdisc}) for $m\ll\mtip$ and $m\gg\mtip$. 

The inclusion of the new parameter $\sigtip$ is motivated by a correct modelling of the observed non-vertical luminosity drop around the tip magnitude. There are several factors that can contribute to keep the transition a smooth function.  Examples on the observational side are provided by underestimation of photometric errors and/or by the unaccounted influence of blending, both of which are expected, however, to have a minimal effect in the cases considered here. From a more physical point of view, it must be kept in mind that the tip is { expected} to { manifest} as a vertical drop in the LF for a stellar population whose stars all share the same age, metallicity, and helium abundance. However, in practical terms, within galaxies like the ones under consideration, these parameters exhibit distributions (with correlations), that should consequently blur the actual LF cut-off corresponding to the RGB tip. Moreover, close to the tip, RGB stars may display small-amplitude variability \citep{wray04}, and their random-phase sampling could also contribute to tip smearing. 
{ Another significant source of broadening of the RGB tip is the possibility that the reference stellar system has a three-dimensional structure and thus a non-negligible depth. At the same luminosity, stars at the tip that are closer (further) to the observer will have a lower (higher) relative magnitude. Consequently, the overall effect of combining stars along the same line of sight but at different depths is to produce a broadening of the RGB tip. This aspect is discussed in detail in Appendix~\ref{app:geo}.}

\begin{figure*}
    \centering
    \includegraphics[width=1\hsize]{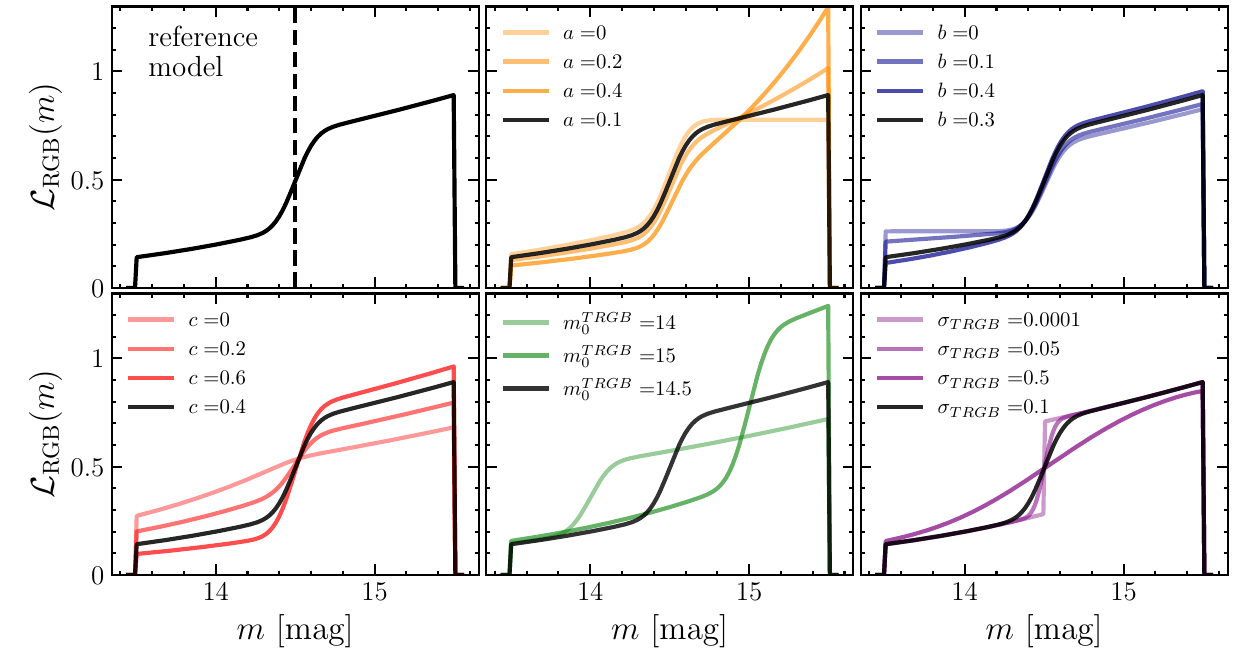}
    \caption{The adopted model of the upper RGB LF~(\ref{for:final}) for different values of its free parameters. The top left panel shows a reference model (black solid line) with $(a,b,c,\mtip,\sigtip)=(0.1,0.3,0.4,14.5,0.1)$ alongside the nominal value of the tip magnitude $\mtip$ (vertical dashed line). The other panels show how the model changes when changing a single parameter while keeping the others to the values of the reference model.}
    \label{fig:prop}
\end{figure*}

To summarise, the five model free parameters are:
\begin{itemize}
    \item $a$: the power-law index for $m\gg\mtip$;
    \item $b$: the power-law index for $m\ll\mtip$;
    \item $c$: a measure for the discontinuity jump around the tip magnitude;
    \item $\mtip$: the magnitude of the tip;
    \item $\sigtip$: the degree of smoothing of the LF around the tip magnitude.
\end{itemize}
In Fig.~\ref{fig:prop} we give a quantitative idea of the main role of each free parameter of model~(\ref{for:final}).

\subsection{The method}
\label{sec:method}

Provided an observed set of $N$ RGB stars sampling the RGB tip transition in a given photometric band with known magnitude, $\mi$, and error, $\delta\mi$, with $i=1,...,N$, we fit the data with model~(\ref{for:final}), thus estimating all its parameters including $m_0^{TRGB}$. We employ a star-by-star fitting technique, avoiding any arbitrary binning of the data. Thus, the model likelihood is given by
\begin{equation}
    \mathscr{L} = \prod_{i=1}^N(\LLRGB\circledast G)(\mi),
\end{equation}
where $G$ is a Gaussian of null mean and dispersion equal to $\delta\mi$. Therefore, each star is included in the fit alongside its photometric error

We conducted a Markov Chain Monte Carlo (MCMC) analysis to explore the parameter space and to sample from the posterior distribution. We adopt  uniform priors on the model free parameters and, to sample from the posterior, we combine the differential evolution proposal by \cite{Nelson2014} and the snooker proposal by \cite{terBraak2008}, relying on the \textsc{emcee} software library \citep{ForemanMackey2013}. In each run we use 30 walkers, each evolved for 5000 steps. We discard an initial burn-in phase of, at least, 500 steps and implement a thinning  of approximately 20 steps, aligning with the auto-correlation length of the chains. The remaining steps are used to build the posterior distributions over the model free parameters. The $1\sigma$ confidence intervals over the models parameters and any derived quantity are computed as the 16th, 50th and 84th percentiles of the corresponding distributions, while the $3\sigma$ confidence intervals as the 0.15th, 50th and 99.85th percentiles.

Further details on the method and tests we performed to evaluate the accuracy of the algorithm are report in appendix~\ref{app:corner}.

\section{Summary and conclusions}
\label{sec:conclu}

We used Gaia DR3 synthetic photometry to assemble large selected samples of LMC and SMC stars with photometry in twenty-two optical passbands of widely used (or to be widely used in the near future) photometric systems, namely Johnson-Kron-Cousins, SDSS, PS1, HST ACS-WFC, JWST NIRCAM, NGRST and Euclid. Having properly corrected all magnitudes for interstellar extinction, we used these samples,  to measure the RGB tip in both galaxies in eight suitable passbands in these photometric systems, using a fully bayesian approach. 

We couple these measures with the state-of-the-art distance estimates for LMC and SMC from eclipsing binaries \citep{lmc19,smc20} to provide calibrations of the RGB tip as a function of colour for the following eight passbands: JKC I, HST ACS/WFC F814W, SDSS i and z, PS1 y, JWST-NIRCAM F090W, NGRST Z087, and Euclid-VIS I$_E$. The calibration in JKC I band was used for a critical comparison with other calibrations in the literature. We find that our calibration is fainter by $\simeq 0.06$~mag than that by H23 and other authors, while it is in good agreement with, e.g., \citet{soltis21}. In Appendix~\ref{app:diff_hoyt} we discuss in some detail the reason of the difference with H23, highlighting the effect of a few factors that may increase uncertainties in $m_0^{TRGB}$ and that are not always taken into account in the literature, including, e.g., systematics in photometric zero points. The latter are expected to affect distance estimates obtained with all standard candles, not only the RGB tip.

The behaviour of the absolute magnitude of the RGB tip as a function of colour (and age) in the various considered passbands is explored also outside the colour range spanned by the LMC and SMC tips by means of theoretical models. SDSS z, JWST NIRCAM F090W, and NGRST Z087 appear as the passbands where the absolute magnitude of the RGB tip displays the lowest dependency on colour, over wide colour ranges. The amplitude of the colour dependencies in these bands is similar (or slightly smaller, in some case, for old age models) to that in JKC I. 

As far as we know, we provide here the first calibrations of the RGB tip in SDSS i and z, PS1 y, NGRST Z087, and Euclid-VIS I$_E$ ever appeared in the literature. An independent calibration in  JWST NIRCAM F090W  has been very recently published by \citet{anand24}, in agreement with our result { within the uncertainties}. Calibrations of the RGB tip analogous to those provided here can be obtained also for additional passbands and as a function of different colours with respect to those we adopted in the present analysis by using our LMC and SMC catalogues that we make publicly available.


\begin{acknowledgements}

We are very grateful to Taylor Hoyt for providing us the material for a thorough comparison with his results and for insightful discussion. We acknowledge the help of Francesca De Angeli with GaiaXPy and Paolo Montegriffo for computing Euclid-VIS I$_E$ XPSP basis functions in the AB system.
MB acknowledges the support to activities related to the ESA/\Gaia mission by the Italian Space Agency (ASI) through contract 2018-24-HH.0 and its addendum 2018-24-HH.1-2022 to the National Institute for Astrophysics (INAF). M.B. acknowledges the financial support by the Italian MUR through the grant PRIN 2022LLP8TK\_001 assigned to the project LEGO – Reconstructing the building blocks of the Galaxy by chemical tagging (P.I. A. Mucciarelli), funded by the European Union – NextGenerationEU.
RP acknowledge the financial support to this research by the Italian Research Center on High Performance Computing Big Data and Quantum
Computing (ICSC), project funded by European Union - NextGenerationEU - and National Recovery and Resilience Plan (NRRP) - Mission 4 Component 2 within the activities of Spoke 3 (Astrophysics and Cosmos Observations).

This work has made use of data from the European Space Agency (ESA) mission \Gaia (https://www.cosmos.esa.int/Gaia), processed by the \Gaia Data Processing and Analysis Consortium (DPAC, https://www.cosmos.esa.int/web/Gaia/dpac/consortium). Funding for the DPAC has been provided by national institutions, in particular the institutions participating in the \Gaia Multilateral Agreement.

In this analysis we made use of TOPCAT (http://www.starlink.ac.uk/topcat/, \citealt{Taylor2005}).

\end{acknowledgements}


\bibliographystyle{aa} 
\bibliography{refs} 


\begin{appendix}

\section{Detection of the RGB tip in different passbands}
\label{app:dete}

\begin{figure}[ht!]
\center{
\includegraphics[width=\columnwidth]{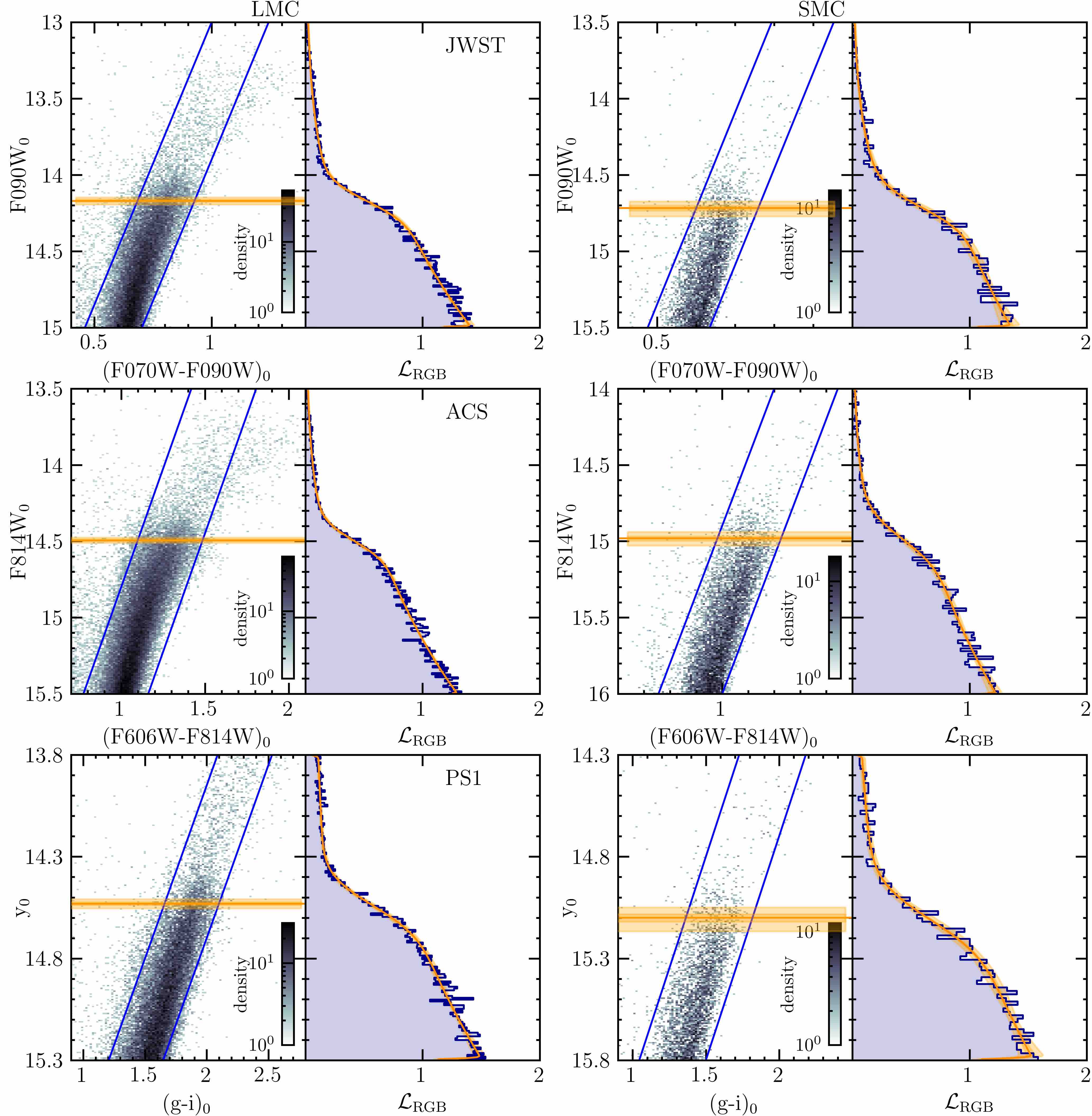}
}
\caption{Detection of the TRGB with our method for the LMC (left column of panels) and the SMC (right column of panels), for, from the upper to the lower row of panels,
JWST-NIRCAM F090W, ACS-WFC F814W, and y$_{PS1}$ passbands, as a function of a colour in the respective photometric systems. The arrangement and the meaning of the symbols are the same as in Fig.~\ref{fig:vitip}.}
\label{fig:rem1}
\end{figure}

\begin{figure}[ht!]
\center{
\includegraphics[width=\columnwidth]{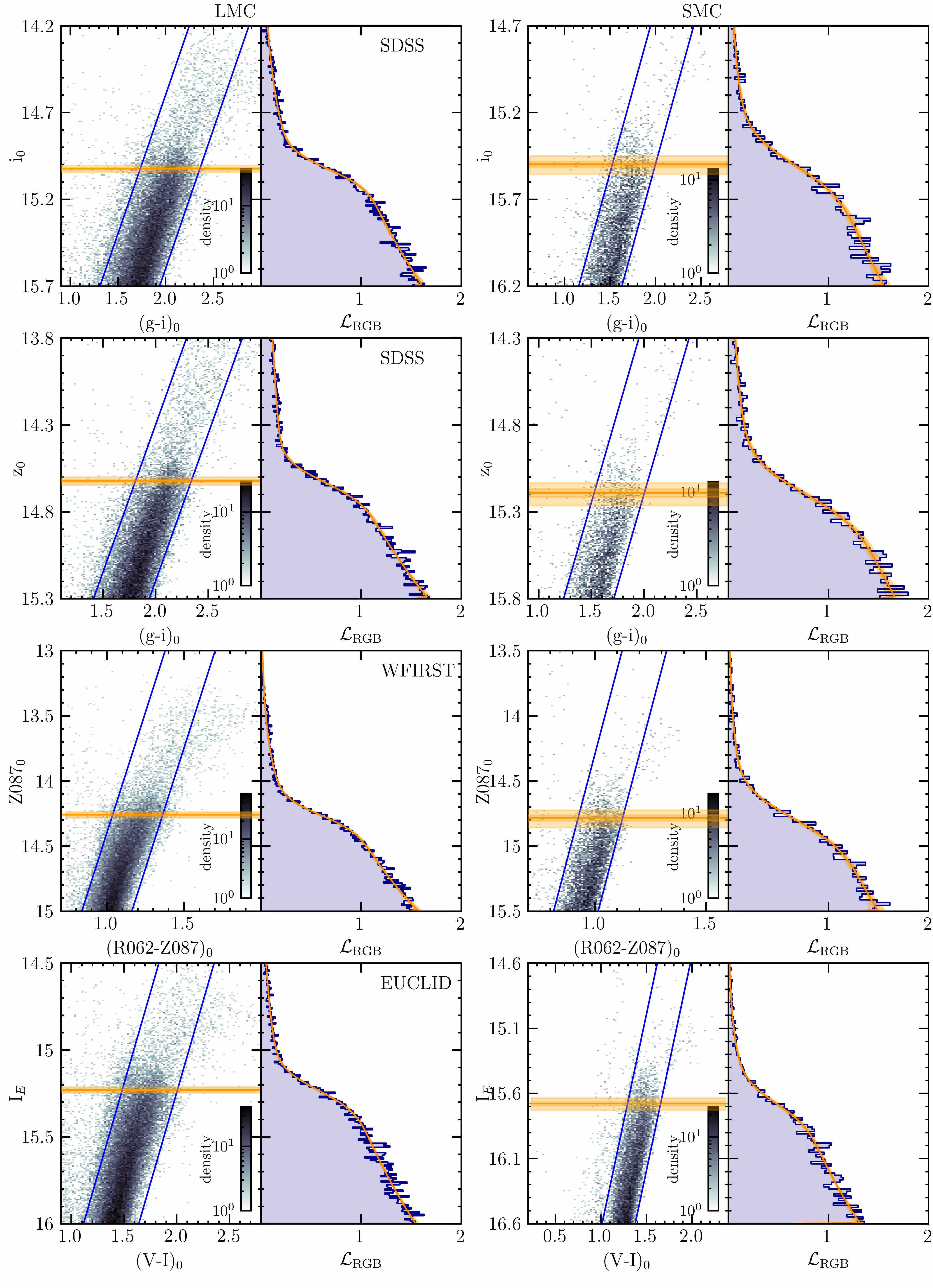}
}
\caption{Detection of the TRGB with our method for the LMC (left column of panels) and the SMC (right column of panels), for, from the upper to the lower row of panels,
SDSS i, SDSS z, NGRST (WFIRST) Z087 and Euclid-VIS I$_E$ passbands, as a function of a colour in the respective photometric systems (except for Euclid, as I$_E$ is the only passband of the system that is completely enclosed in the spectral range of Gaia XP spectra; in this case we used the V-I colour). The arrangement and the meaning of the symbols are the same as in Fig.~\ref{fig:vitip}.}
\label{fig:rem2}
\end{figure}

In Fig.~\ref{fig:rem1} and Fig.~\ref{fig:rem2} the detection of the RGB tip in the SMC and LMC are shown for the various passbands considered here in addition to JKC I
(see Sect.~\ref{sec:measure}). The arrangement and the meaning of the symbols are the same as in Fig.~\ref{fig:vitip}. The detection is fairly clean in all the considered cases.

\section{The origin of the differences with H23 and of the excellent match with Y99}
\label{app:diff_hoyt}

In Sect.~\ref{sec:measure} we showed that our measures of the extinction-corrected
apparent magnitudes of the JKC I band RGB tip of the LMC and SMC are $\simeq 0.06$~mag fainter than the corresponding state-of-the-art measures by H23. In this appendix we try to understand the reason for this difference by exploring a few possibly promising hypotheses. 

\begin{figure}[ht!]
\includegraphics[width=1\hsize]{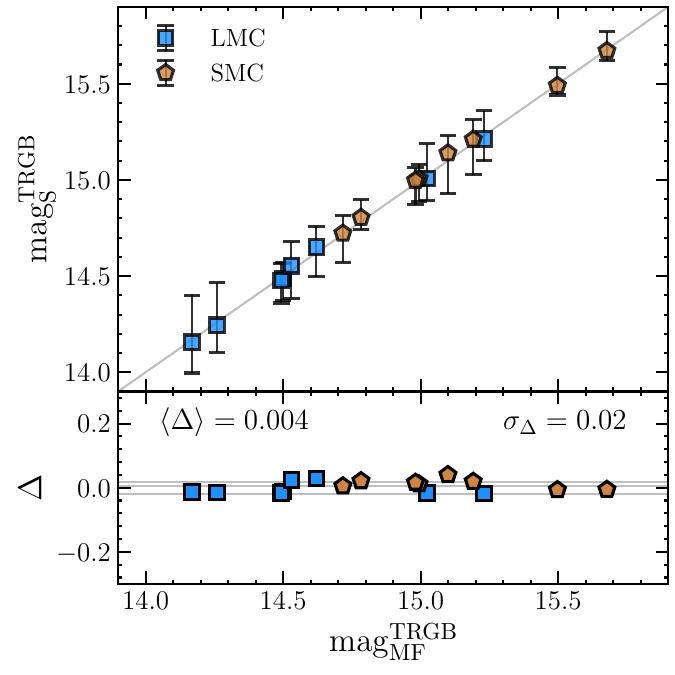}
\caption{Upper panel: comparison between the magnitudes of the TRGB in the LMC (blue filled squares) and in the SMC (orange filled pentagons) in all the considered passbands measured with the MF algorithm  (mag$_{MF}^{TRGB}$) and those mesured with the Sobel filter (mag$_{S}^{TRGB}$). The thick grey line is the bisector of the panel. Lower panel: differences between the two values as a function of mag$_{MF}^{TRGB}$. The mean and the standard deviation of the differences is also reported. { The horizontal grey lines mark the mean and the $\pm 1\sigma$ range about the mean}.}
\label{fig:sobML}
\end{figure} 

First, it has to be noted that in H23 the detection of the RGB tip is obtained with the edge-detector filter technique while here we adopted the model fitting (MF) method described in Sect.~\ref{sec:method}. In principle, there is no guarantee that the two methods locate the tip exactly at the same position \citep{anand22,anand24}. To test this possibility we repeated all the RGB tip detections presented in Sect.~\ref{sec:measure} and Appendix~\ref{app:dete} using the Sobel filter technique \citep{sakai96} exactly on the same samples of colour selected RGB and AGB stars. In all the cases the TRGB was unambiguously identified as a dominant peak in the filter response. We assumed the half width at half maximum (HWHM) of the filter response peak associated to the tip as the uncertainty on the magnitude of the tip. This is probably an overestimate of the actual uncertainty but it is a simple and well defined quantity. Moreover, here we are mainly interested to check if there is a systematic difference between the two methods in the actual location of the tip. The upper panel of Fig.~\ref{fig:sobML} shows the comparison between the magnitude of the tip obtained from the two different techniques for the sixteen cases considered in this paper (8 passbands $\times$ 2 samples). All the points lie very close to the mag$^{TRGB}_{S} = $mag$^{TRGB}_{MF}$ line and the residuals (mag$_S^{TRGB}$-mag$_{MF}^{TRGB}$), plotted in the lower panel of the figure, are strongly clustered around zero, with a mean as small as +0.004~mag and a standard deviation of 0.018~mag, with no obvious sign of systematics. From this experiment we can conclude that it is very unlikely that the detection method is responsible for the difference between our estimates of the MC tips and those by H23, but also that there is an unavoidable amount statistic uncertainty intrinsically associated to the way in which the measure is performed, here of order of $0.02$~mag \citep{anand22}. The RGB tip is a somehow anomalous standard candle, where the reference observable is not, e.g.,  the mean magnitude of any individual source of a given class but, instead, a collective property of a stellar population that must be inferred, hence it is not surprising that the inference method can play a role in the reproducibility of measures. It is important to be aware that this limit in accuracy is inherent to each measure of the tip even if it is generally neglected\footnote{However, in principle, the effects of this limit in precision can be mitigated within a programme where all the tip measures are performed with the same technique and with the same selection criteria, from the calibrating sources to the farthest target galaxy.}. 
In the specific case of the the JKC I band for the LMC the measure obtained with the Sobel filter { is $0.012$~mag} brighter than that obtained with the MF algorithm. 
Hence, { a small part} of the difference between us and H23 in $I_0^{TRGB}$ for this galaxy can be ascribed to a $\simeq 1\sigma$ fluctuation in this intrinsic term of the uncertainty budget.

\begin{figure}[ht!]
\includegraphics[width=\columnwidth]{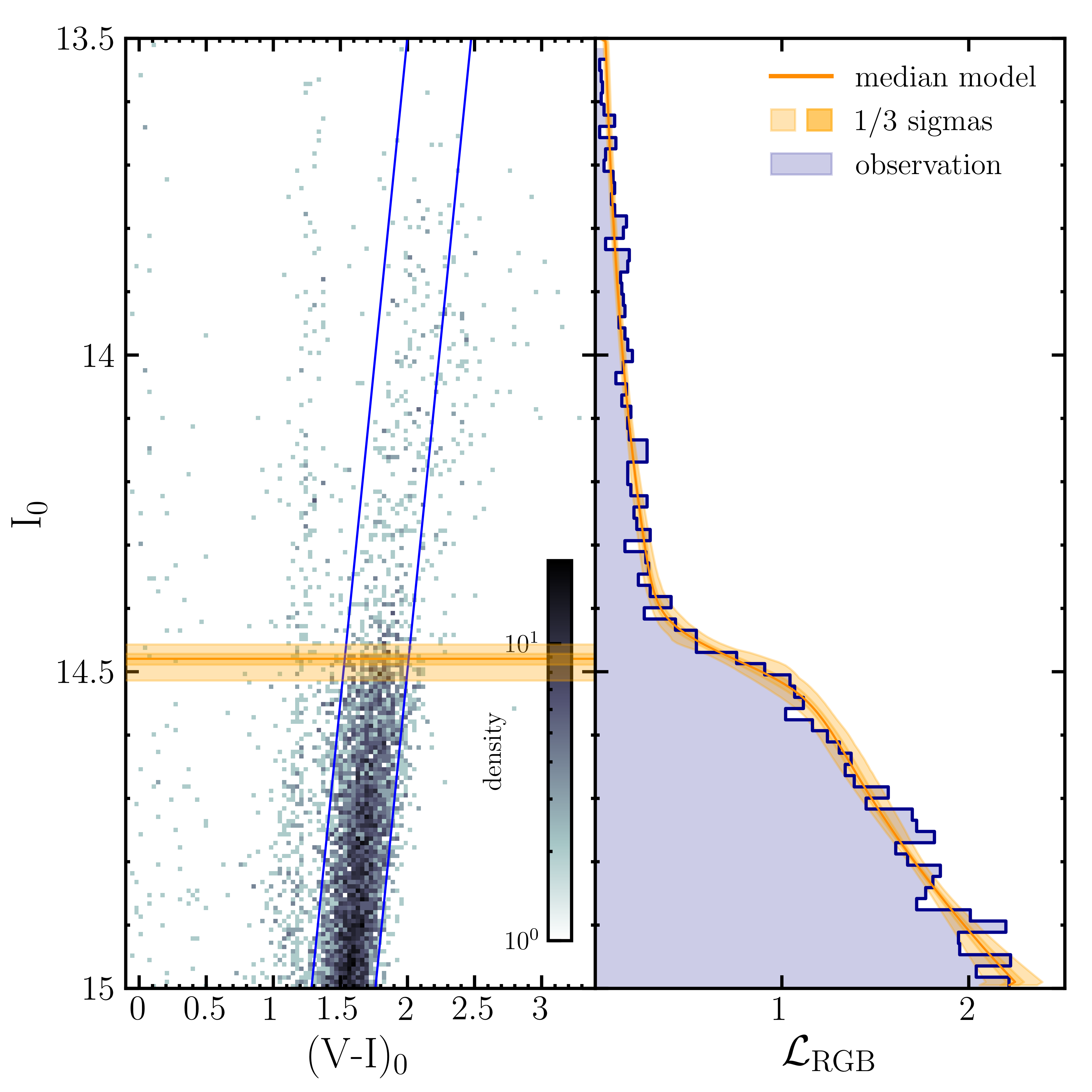}
\caption{Detection of the RGB tip in the JKC I band for the LMC adopting only stars lying within the Voronoi bins used by H23 for his calibration. 
The arrangement and the meaning of the symbols are the same as in Fig.~\ref{fig:vitip}. }
\label{fig:hoytbins}
\end{figure} 

A second issue that we explore is the detailed selection of the sample adopted for the detection, in particular for the LMC. H23 used the Voronoi tessellation of the OGLE-III sample obtained in \citet{hoyt18} and made a detailed analysis to select the bins ensuring the most precise detection of the LMC tip. To check if differences in the adopted sample is at the origin of the differences in the tip measures we repeated the tip detection for the LMC using the stars in our sample enclosed in the same five Voronoi bins selected by H23. Moreover we adopted the same correction for the inclination of the LMC disc used by H23\footnote{We note however that the measures of the tip we obtained from this sample with and without the correction for disc inclination are statistically indistinguishable.}. The detection of the tip with this sub-sample is shown in Fig.~\ref{fig:hoytbins}. We confirm that the detection in this sample is especially clean and more precise that in the entire sample we used elsewhere. For instance, the HWHM of the peak in the Sobel filter response is $\pm 0.040$~mag for the tip detection with this subsample and $_{-0.011}^{+0.084}$ for the detection with the entire sample. 

The actual measure of the tip with the MF method for the H23 selected sample is $I_0^{TRGB}=14.480_{-0.008}^{+0.010}$ to be compared to $I_0^{TRGB}=14.500_{-0.006}^{+0.007}$ for the entire sample. { The difference is $<3\sigma$ but it goes in the direction of reconciling our measure with that by H23.} Still, the amplitude of the effect (0.02~mag) is not sufficient to fill the observed gap between the two measures (0.06~mag). 

\begin{figure}[ht!]
\includegraphics[width=\columnwidth]{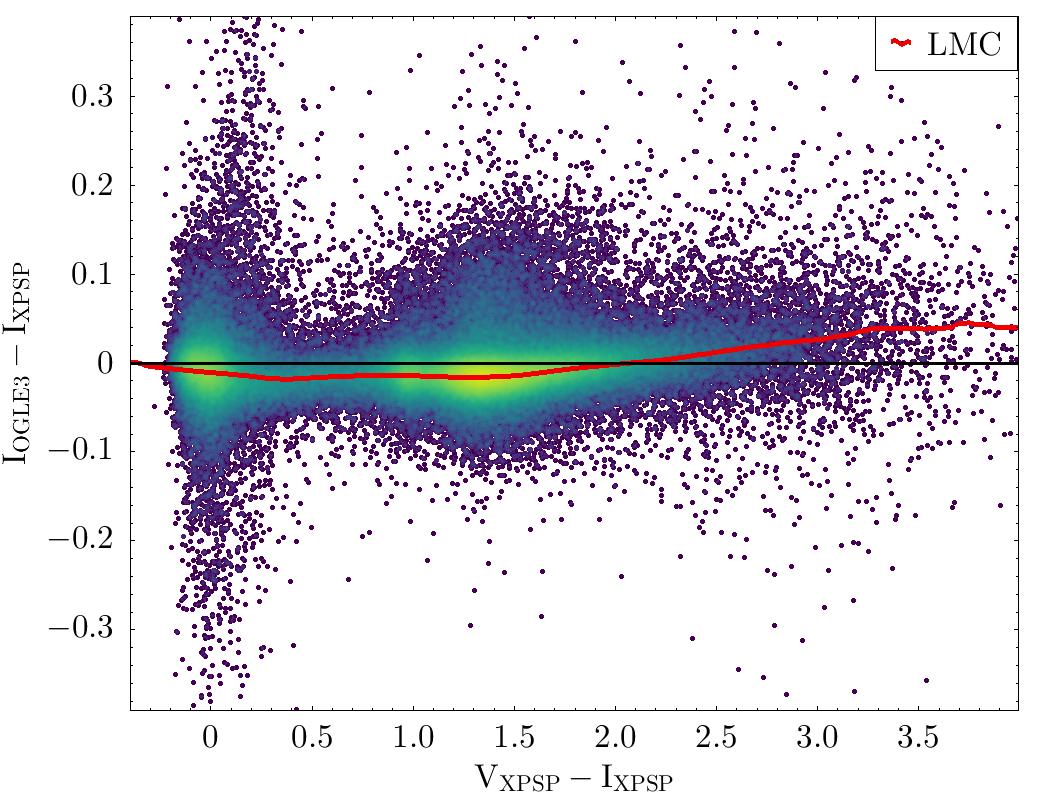}
\includegraphics[width=\columnwidth]{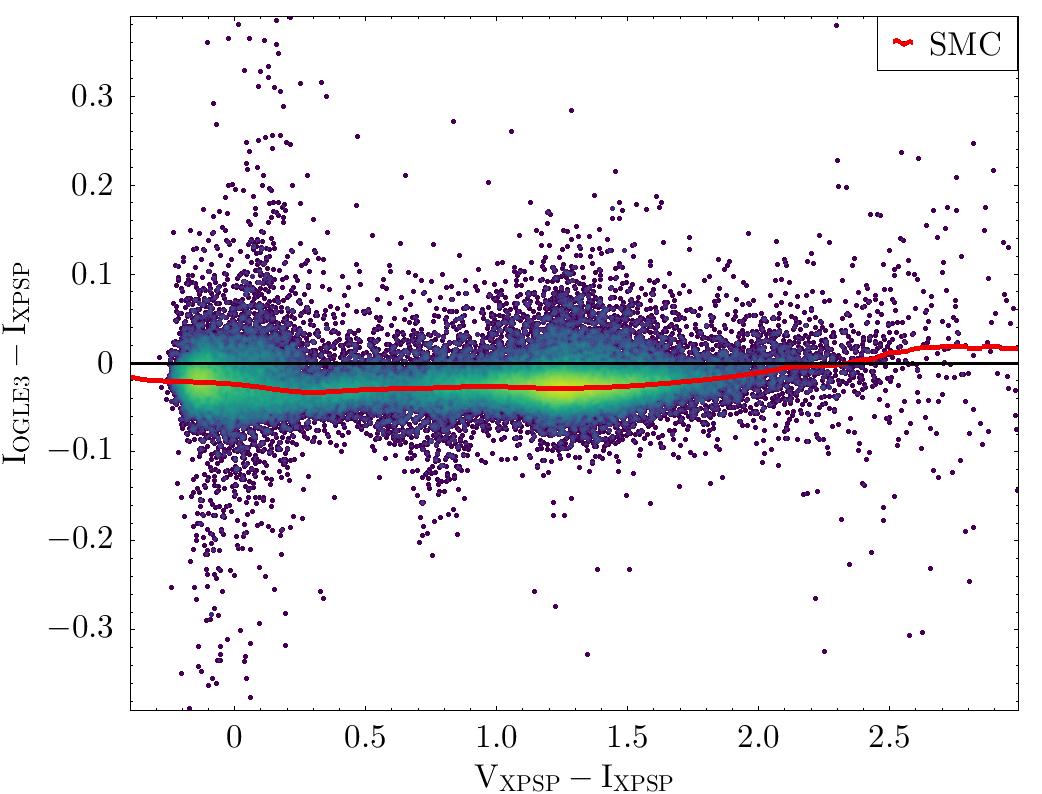}
\caption{Difference between I magnitudes from our XPSP samples and OGLE-III for
315080 stars in common between the two samples having $|\Delta I|<1.0$ (LMC; upper panel) and 90115 stars in common between the two samples having $|\Delta I|<1.0$ (SMC; lower panel). In each panel, the red line is the median difference computed in bins 0.2~mag wide.}
\label{fig:di_vi}
\end{figure} 

\begin{figure}[ht!]
\includegraphics[width=\columnwidth]{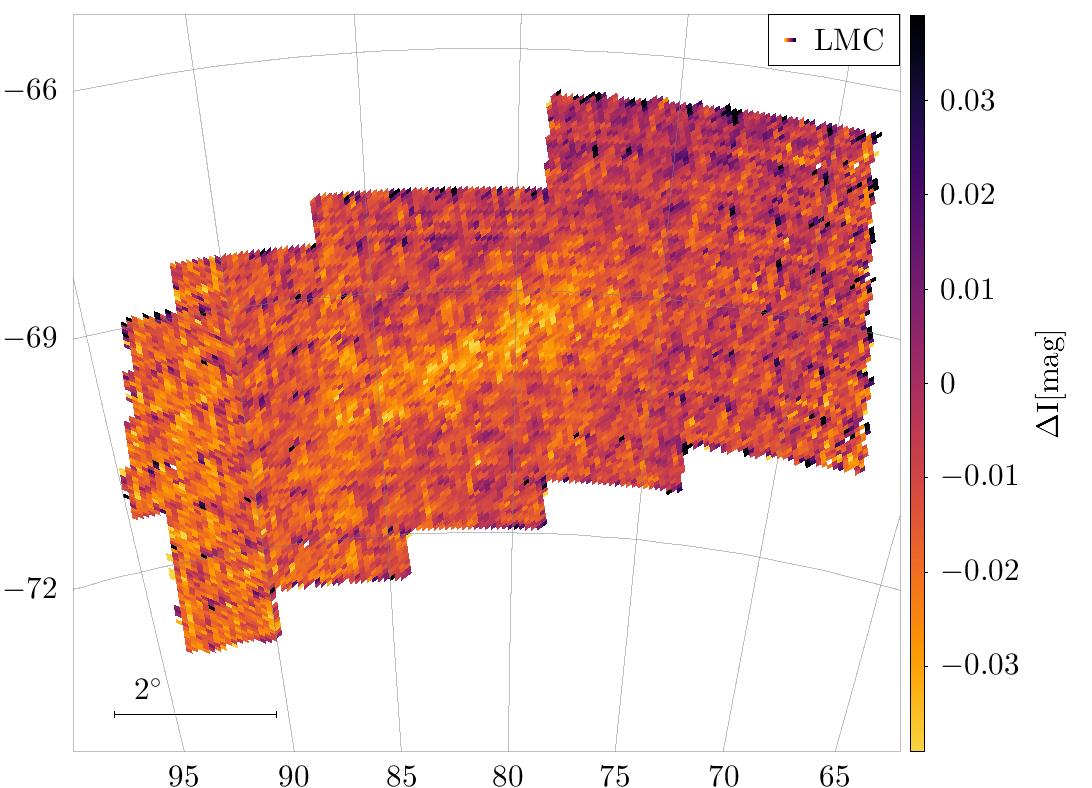}
\caption{LMC healpix map (HEALPiX level 10) obtained from the same subsample of stars in common between out XPSP sample and the OGLE~III sample adopted in Fig.~\ref{fig:di_vi}. The pixels are coloured according the the median difference $I_{OGLE3}-I_{XPSP}$.}
\label{fig:di_map}
\end{figure}

Finally, we compared the photometric zero points. In Fig.~\ref{fig:di_vi} we show the distribution of the difference in I magnitudes between OGLE-III photometry and our XP standardised synthetic photometry for the stars in common between the two datasets in the LMC and SMC, as a function of V-I colour. In both cases, OGLE-III I band magnitudes display a photometric zero point brighter by 0.02-0.03~mag than that of our Gaia XPSP sample. Also this difference goes in the direction of reconciling the two sets of RGB tip measures, in this case by correcting H23 measures toward fainter magnitudes. Also in this case, the amplitude of the effect does not seem sufficient to close the gap. It is worth noting that the general trends with colour shown in Fig.~\ref{fig:di_vi} hide trends with position in the sky that are mapped in Fig.~\ref{fig:di_map} for the LMC, as an example. In this plot each healpix
is coloured according to the median of the distribution of the $\Delta I = I_{OGLE3}-I_{XPSP}$ difference for the stars in the healpix, after removing from the sample all the extreme outliers having $\Delta I>1.0$~mag. Both large and small scale trends with typical amplitude
$\la 0.04$~mag are apparent. The fact that they correlate clearly with the tiling pattern of OGLE-III strongly suggest that XPSP is the most precise source of photometry in this comparison. 

Indeed \citet{dr3_dpacp93} showed that XPSP reproduces the Landolt's incarnation of the JKC system in I band with a typical accuracy $<3.0$~mmag and precision of $\simeq 10$~mmag, in the range $11.0<G<17.5$. Moreover XPSP photometry  displays a remarkable uniformity all over the sky \citep[see, e.g.,][]{dr3_xp_paolo}, and is being widely used as a reference to obtain more consistent photometric calibrations of existing surveys \citep[see, e.g.,][]{martin23,xiao23a,lopez23}.

This last comparison highlights a general issue that may affect all standard candles: state-of-the-art photometric datasets (in particular ground-based ones) may have inconsistencies in the photometric zero points at the 2-3\% level, including trends with colour and/or position in the sky. The effects of this generally neglected source of uncertainty can be significantly mitigated, at least for the photometric systems standardised by \citet{dr3_dpacp93}, by the availability of the all-sky
Gaia XPSP, providing, in practice, a reliable and dense grid of standard stars to which all photometric observations can be referred. 

\subsection{Comparison with Yuan et al. (2019)}
{

In this section we briefly illustrate how the same kind of difference in the photometric zero point that mitigates the mismatch between our calibration of $M_I^{TRBG}$ for the LMC with that by H93, may act to worsen the (apparently) nearly perfect match with that by Y19, who finds $M_I^{TRBG}=-3.970\pm 0.046$, formally within 0.007~mag of our value. In Fig.~\ref{fig:y19comp} we show that the Y99 F814W photometric scale is $\simeq 0.03$~mag brighter than our XPSP in the same passband, approximately corresponding to $\simeq 0.04$~mag in $I_{JKC}$ band. This translates into a difference in the same sense but slightly larger than that found with OGLE3, in good agreement with the comparison between OGLE3 and Y19 photometry presented in Y19. 

\begin{figure}[ht!]
\includegraphics[width=\columnwidth]{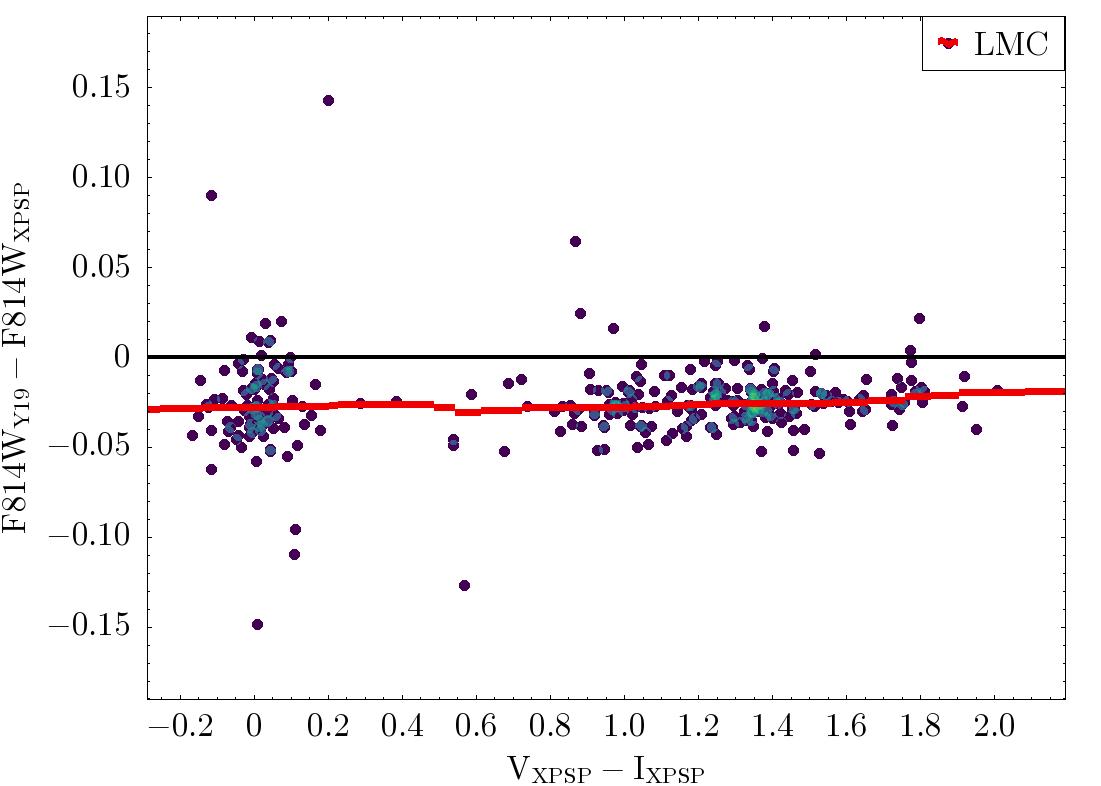}
\caption{Difference between F814W magnitudes from our XPSP LMC sample and Y19 for
316 stars in common, as a function of V-I colour. The red line is the median difference computed in bins 0.3~mag wide. The vertical scale and the overall arrangment of the plot are the same as in Fig.~\ref{fig:di_vi} to allow a direct comparison.}
\label{fig:y19comp}
\end{figure} 

If we correct the Y19 $M_I^{TRBG}$ value by this photometric shift, the difference between our LMC calibration and theirs inflates to $-0.05\pm 0.06$, still not significant but much larger than comparing only the final calibrated values. This exercise may serve as an additional example of the possible role of the uncertainty in the photometric zero points in the use of standard candles for distance determination.

The photometric zero point difference that we detect here between Gaia XPSP and Y99 photometry in  F814W probably implies a difference in zero point between Y99 and the reference photometry we used to standardise XPSP in the ACS/WFC system \citep{nardiello18}\footnote{See \citet{dr3_dpacp93} for further details on this specific XPSP standardisation.}. 
This should not come as a strong surprise, since, while the photometric precision achievable with HST is tipically exquiste, the calibration of individual photometric sets into the absolute scale may be subject to various sources of uncertainty. For example, according to the detailed study by \citet{bedin05}, the simple correction of the conventional $r=0.5\arcsec$ aperture to infinity is associated to an uncertainty as large as $0.015$~mag. To account for this factor and given the results of the comparison presented here, in the error budget that includes the uncertainty in the photometric zero point in Table~\ref{tab:tipcal} we adopted a 3\% additional uncertainty on the $M_{F814W}^{TRBG}$ calibration.

}

\section{Selection and geometric effects}
\label{app:geo}

{

\begin{figure*}[ht!]
\includegraphics[width=\textwidth]{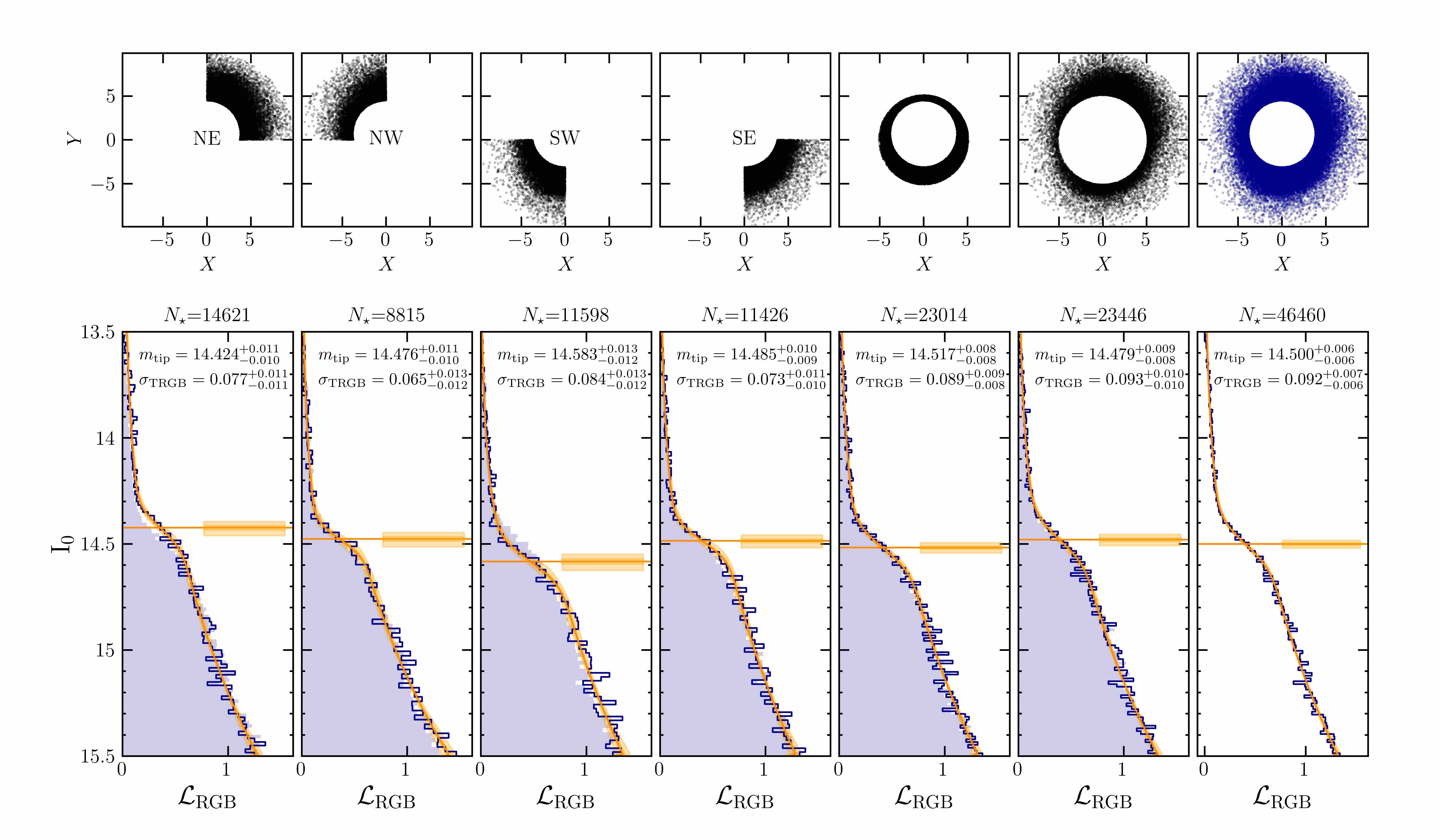}
\includegraphics[width=\textwidth]{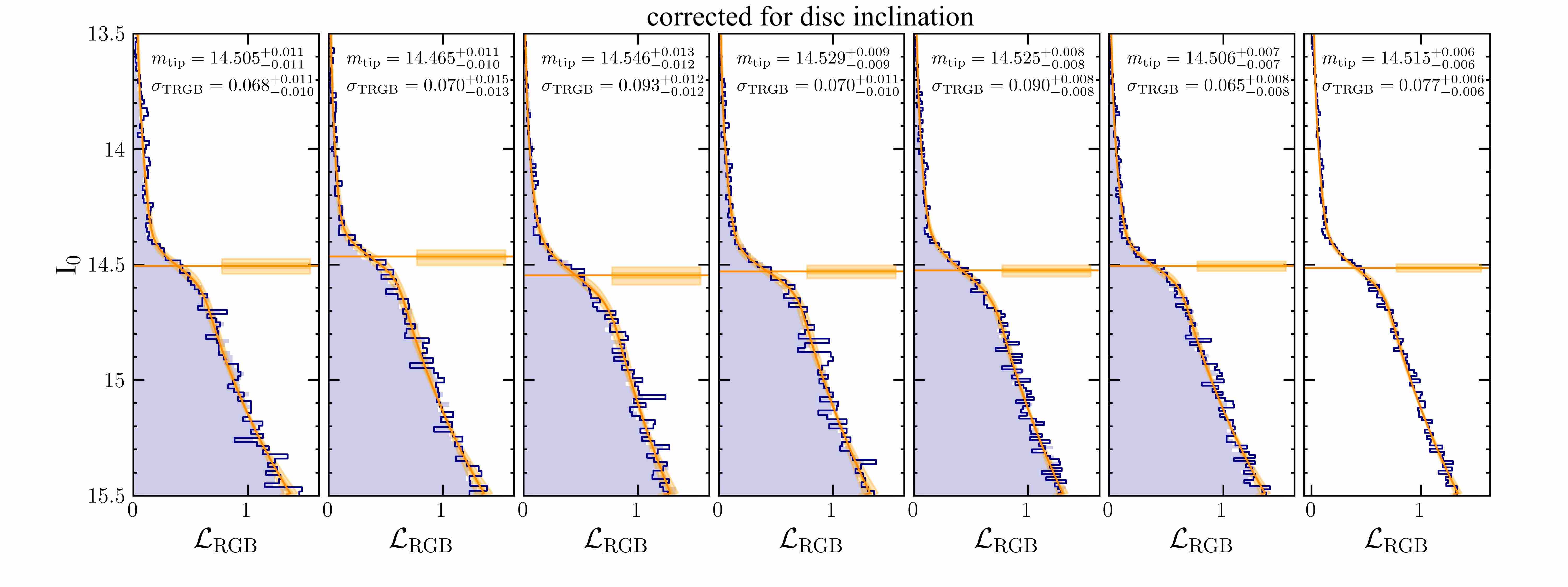}
\caption{Upper row of panels: maps of the stars in the different subsamples of the LMC samples used for the RGB Tip detections shown in the lower panels. The inclination of the LMC disc is such that the NE side is the closest to us and the SW side is the farthest. Middle row of panels: Detection of the RGB Tip in different quadrants (four leftmost panels) and different radial zones (fifth and sixth panels from the left) of the LMC sample. In the eighth panel the detection on the entire sample is shown.
Lower row of panels: the same as in the middle raw but after applying the correction for the 3d geometry of the LMC.
}
\label{fig:quad}
\end{figure*}

One drawback of using the Magellanic Clouds as the basis for the calibration of standard candles is that both galaxies display a 3d structure that, given their relatively closeness to us, has non-negligible impact on the apparent magnitude of the candles. The main body of LMC is a thick disc with relatively low  inclination \citep[$i\simeq 33\degr$;][]{edr3_luri}, such that the NE edge is $\sim 4$~kpc closer to us than the opposite SW edge \citep[see][and references therein]{edr3_luri}. Moreover, it has been suggested that the disc is warped in its outskirts \citep[beyond $R\simeq 7.0\degr$, according to][]{lmc_warp}. On the other hand, the SMC is known to be significantly elongated along the line of sight, but with an irregular structure, probably due to tidal interactions with the LMC \citep[][and references therein]{okumar21,tatton21,zivick21, cullinane2023}. \citet{SMC_is_two} recently presented evidence supporting the idea that what we call SMC is in fact the superposition of two bodies located at slightly different distances along the same line of sight, and displaying different internal kinematics.

A detailed modelling of the 3d structure of the LMC and SMC is clearly beyond the scope of the present contribution. Our approach is to try to average out these effects by the adoption of large samples with an approximately uniform distribution in azimuth and including a parameter ($\sigma_{TRGB}$; Sect.~\ref{sec:method}) in the model used to detect the RGB Tip that can account for any effects that can contribute to smear out the Tip signal, including dispersion due to geometrical 3d structure. However it is worth inspecting the typical size of these effects.

For the LMC, in Fig.~\ref{fig:quad} we show the RGB detection in I$_{JKC}$ on sub-samples from the various quadrants and from an inner and outer radial region. 
X and Y are orthographic projection coordinates in degrees computed according to Eq.~1 by \citet{edr3_luri}.
Each quadrant provide approximately the same contribution, in number of stars, to the entire sample and the inner and outer sub-samples have similar dimension by intentional choice of the dividing radius. Finally we also report the detection obtained for the entire sample, for reference. The analysis described above is made on the LMC sample as it is and also after correcting it for the disc inclination, using Eq.~5 of H23 with the model parameters taken from \citet{edr3_luri}.

The detections in the various quadrants in the original sample clearly show the significant impact of the 3d structure of the galaxy in our sample, that is dominated by stars outside the LMC center, where the effects of the disc orientation along the line of sight are more significant. Indeed, the distance difference inferred from the TRGB tip difference between the SW and NE quadrants is $\simeq 3.7$~kpc. The simple model adopted to correct for the 3d structure of the disc significantly mitigates these difference, reducing the maximum distance difference between quadrants to  $\simeq 1.9$~kpc, within the range of the $3 \sigma$ uncertainty (neglecting the contribution to the error budget of the photometric zero point, as here this is the same for all the detections). The differences between the detections in the inner and outer sub-samples are relatively small in both cases (+0.038~mag in the original sample and +0.019~mag in the corrected sample). However, the most interesting comparison shown in Fig.~\ref{fig:quad} is that between the detections on the entire sample: the difference in I$^{TRGB}_0$ between the original and the corrected samples is a mere $-0.015\pm 0.008$~mag, the two detections having similar uncertainties. This demonstrate that our strategy of averaging out on the 3d structure has been successful, at least in this case. 
It is also interesting to note that $\sigma_{TRGB}$ is 20\% larger in the detection from the original sample than in that in the corrected one, suggesting that this parameter can properly take into account smearing effects like those due to the 3d structure of stellar systems.

Since no simple model is available to attempt a correction for the 3d structure of the SMC, we simply repeat the above analysis for the original SMC sample, to gauge the amplitude of the differences in the measure of the tip as a function of azimuth and angular distance from the centre of the galaxy. The results are displayed in Fig.~\ref{fig:quadSMC}. The comparison between the various quadrants indeed reveals a complex structure both in term of mean distance, with differences up to $+0.158\pm 0.040$~mag in $I_{0}^{TRGB}$ (between the SW and the NE quadrants), and, presumably, in term of depth along the line of sight, with differences in the $\sigma_{TRGB}$ parameter up to $0.114\pm 0.035$~mag (between the NE and the NW quadrants).

\begin{figure*}[ht!]
\includegraphics[width=\textwidth]{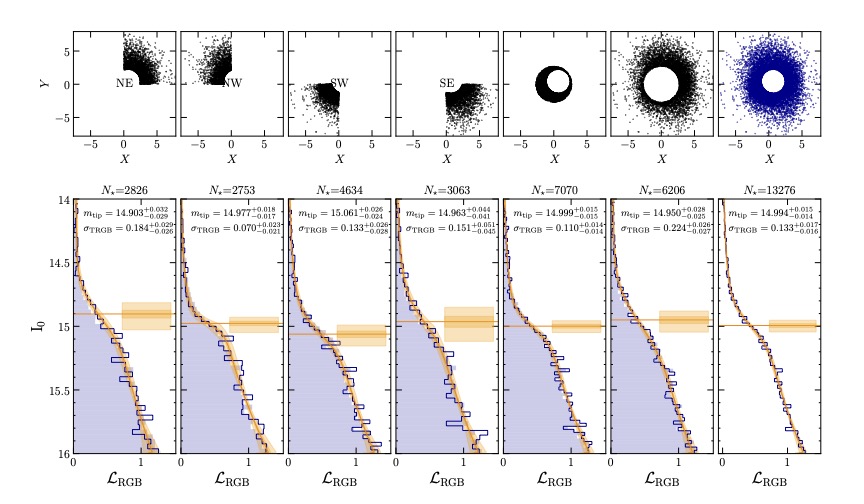}
\caption{Upper row of panels: maps of the stars in the different subsamples of the SMC samples used for the RGB Tip detections shown in the lower panels.  Lower row of panels: Detection of the RGB Tip in different quadrants (four leftmost panels) and different radial zones (fifth and sixth panels from the left) of the LMC sample. In the eighth panel the detection on the entire sample is shown.
}
\label{fig:quadSMC}
\end{figure*} 

The SW quadrant, having the largest $I_{0}^{TRGB}$ value is also the most populated one, while the NE quadrant is the least populated one. It is hard to say if this unbalance may produce a bias with respect to the (unknown) {\em mean} distance of the SMC. However we verified that the mean of the tip measures of the four quadrants weighted by the number of stars, $I_{0}^{TRGB}=14.991$, matches very well with the measure obtained from the entire sample, $I_{0}^{TRGB}=14.994$, hence we can evaluate the impact of the various quadrants on the final mean in a simple way. If we remove the SW quadrant from the weighted mean we obtain a $I_{0}^{TRGB}$ value brighter by 0.037~mag, if we remove the NE quadrant we obtain a $I_{0}^{TRGB}$ value fainter by 0.024~mag. Hence, in spite of the different sample size, the contribution of the two most extreme quadrants compensates in the determination of $I_{0}^{TRGB}$ from the total sample at the $\simeq 0.01$~mag level. The $0.049$~mag difference in $I_{0}^{TRGB}$ between the inner and outer halves of the SMC sample is mainly driven by the asymmetry of the adopted selection. This implies that the inner sample is mainly populated by stars from the SW quadrant, driving the tip to faintest magnitudes with respect to the outer sample. In this context, the approach of averaging out all these differences into a global measure appears as a simple, sensible choice, keeping in mind that the highly irregular 3d structure of the SMC may add some uncertainty to our calibration.

}

\section{The value added LMC and SMC catalogues}
\label{app:cata}

The photometric catalogues on which the present study is based are made publicly available as binary fits tables that can be retrieved from \url{}{\tt [Zenodo link]}, one for the LMC (603311 stars) and one for the SMC (124578 stars). The way in which the stars to be included in the catalogues have been selected as well as how it has been obtained the E(B-V) value associated to each star are described in detail in Sect.~\ref{sec:samples}. 

Each catalogue contains:

\begin{itemize}

\item{} Column 1-15: parameters directly extracted from the Gaia DR3 source catalogue\footnote{See \url{https://gea.esac.esa.int/archive/} and \url{https://www.cosmos.esa.int/web/gaia-users/archive/gdr3-documentation}}, namely: {\tt source\_id}, {\tt ra, dec}, {\tt parallax, parallax\_error}, {\tt pmra, pmra\_error}, {\tt pmdec, pmdec\_error}, {\tt ruwe}, {\tt phot\_variable\_flag}, and {\tt non\_single\_star}.  

\item{} Column 16: $C^{\star}$, a parameter that is derived from {\tt  phot\_bp\_rp\_excess\_factor} (that, in turn, has been extracted from the Gaia DR3 source catalogue) following \citep{edr3_riello21}.

\item{} Column 17: colour excess E(B-V), derived from \citet{skowron21} and \citet{sfd98,schlafly2011} as described in Sect.~\ref{sec:samples}.

\item{} Columns 18-27: standardised ugriz magnitudes in the SDSS system extracted from the Gaia DR3 GSPC \citep{dr3_dpacp93} and associated uncertainties.

\item{} Columns 28-37: standardised UBVRI magnitudes in the JKC system extracted from the Gaia DR3 GSPC \citep{dr3_dpacp93} and associated uncertainties.

\item{} Columns 38-47: standardised grizy magnitudes in the PS1 system computed with GaiaXPy and associated uncertainties.

\item{} Columns 48-51: standardised F606W, F814W magnitudes in the HST ACS-WFC system extracted from the Gaia DR3 GSPC \citep{dr3_dpacp93} and associated uncertainties.

\item{} Columns 52-55: non-standardised F070W and F090W magnitudes in the JWST-NIRCAM system computed with GaiaXPy and associated uncertainties.

\item{} Columns 56-59: non-standardised R062 and Z087 magnitudes in the NGRST system computed with GaiaXPy and associated uncertainties .

\item{} Columns 60-61: non-standardised I$_E$ \citep[][]{cuillandre24} magnitude in the Euclid system computed with GaiaXPy and associated uncertainties \citep[in this case, not corrected for the bias described in Sect. 2.1 of][hence slightly underestimated]{dr3_dpacp93}.
   
\end{itemize}

It is very important to remind that u$_{SDSS}$ and U$_{JKC}$ magnitudes from Gaia XPSP are significantly less accurate and precise than all the other magnitudes included in our catalogues \citep[see][for detailed discussion]{dr3_dpacp93}.
Each star can lack some of the 22 magnitudes listed in the catalogues, depending on the signal-to-noise constraints applied to the original GSPC \citep[S/N$>30$][]{dr3_dpacp93}. For example, of the 603311 stars listed in the LMC sample only 83229(85338) have valid u$_{SDSS}$(U$_{JKC}$) magnitudes.

The catalogues can be useful for several applications, like, for example, to replicate the analysis presented here, to obtain calibrations of the RGB tip in passbands (or as a function of colours) not considered here, for careful selection of targets for spectroscopic follow-up, and to test stellar models in different photometric systems.

\section{Details on the RGB tip fitting procedure}
\label{app:corner}

\begin{figure*}
    \centering
    \includegraphics[width=0.48\hsize]{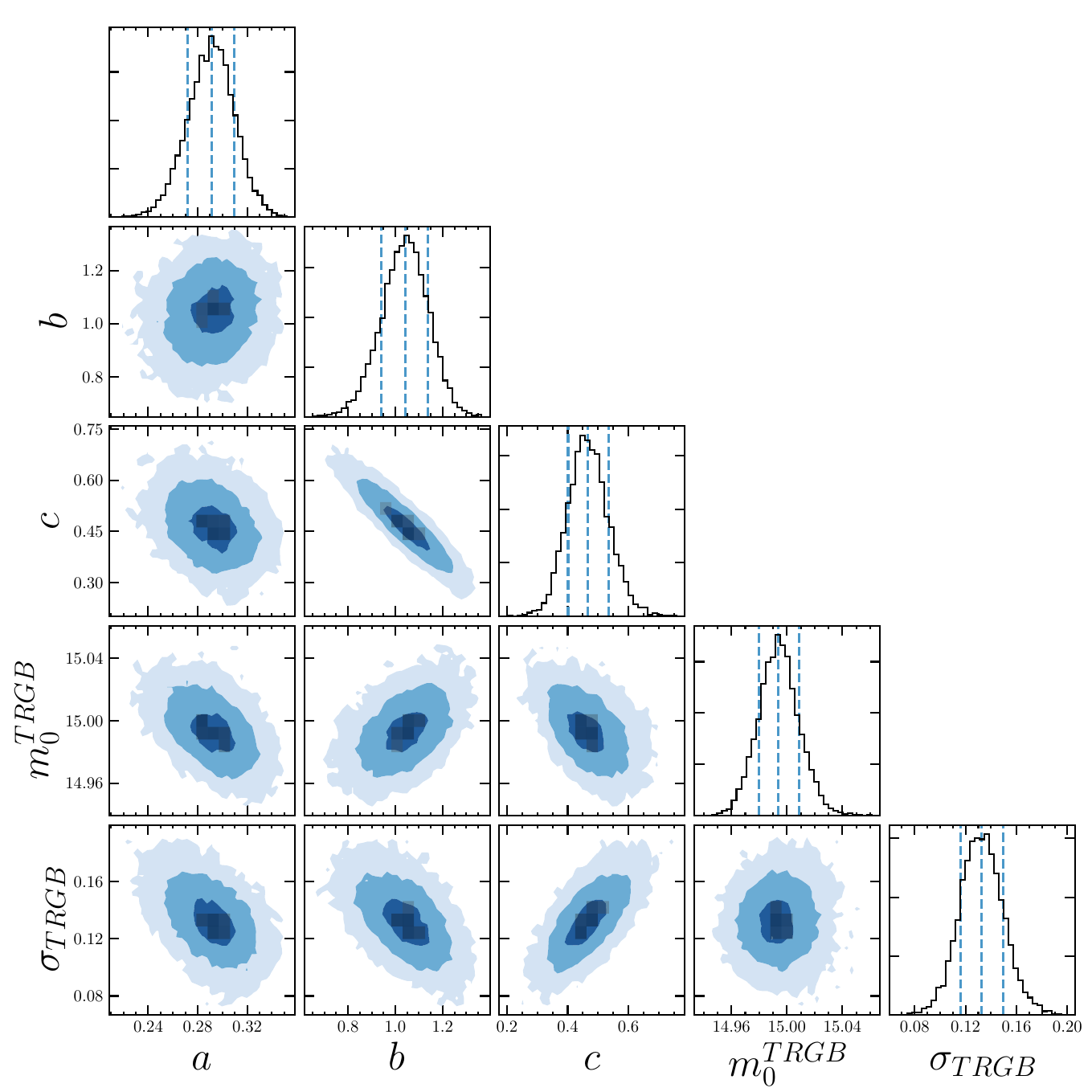}
    \includegraphics[width=0.48\hsize]{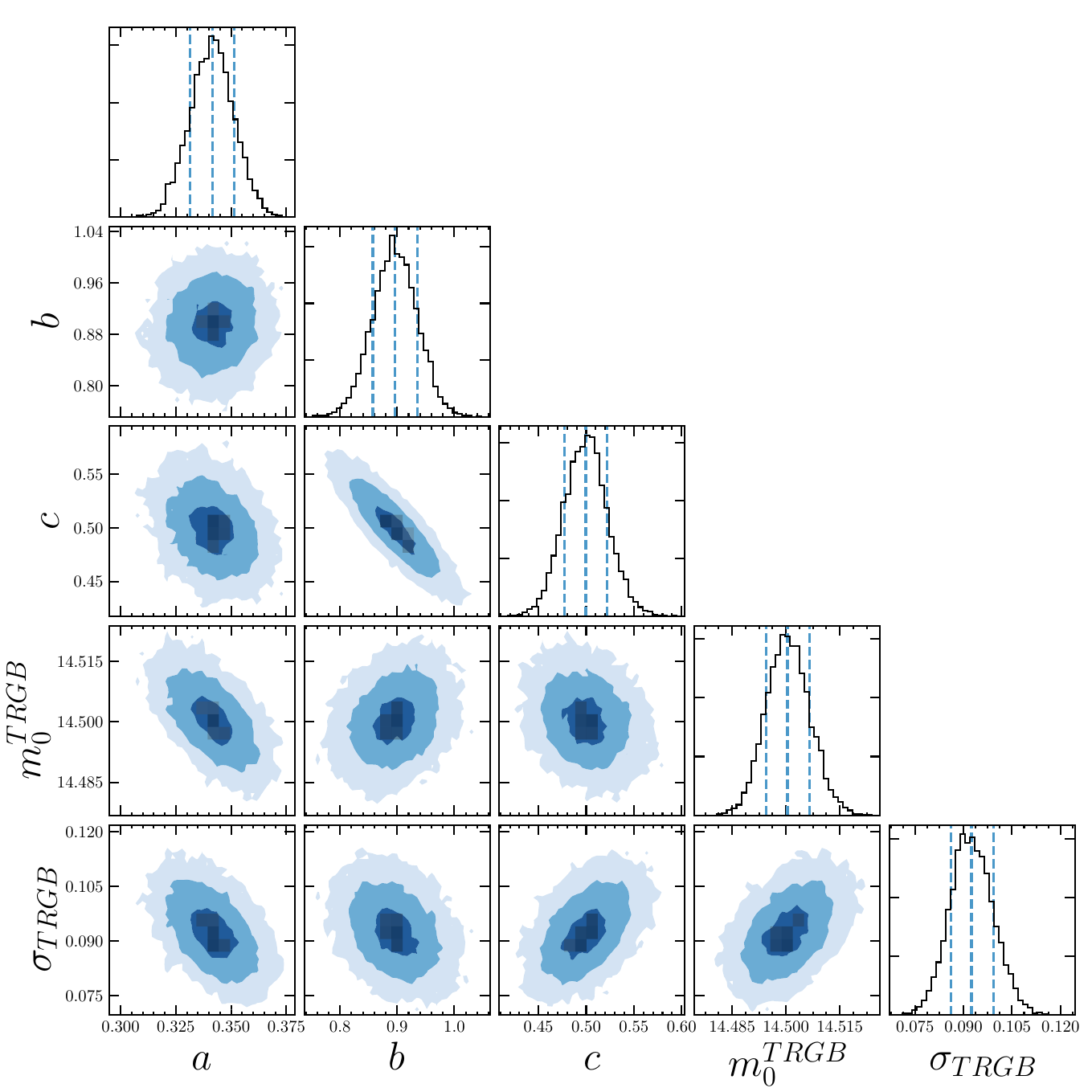}
    \caption{Examples of two- and one-dimensional marginalized posterior distributions on the model free parameters for the SMC (left) and LMC (right) in the JKC passbands (see also Section~\ref{sec:measure}). The vertical dashed-blue lines in the one-dimensional distributions show the 16th, 50th and 84th percentiles used to evaluate the parameters $1\sigma$ confidence intervals.}
    \label{fig:corner}
\end{figure*}

Fig.~\ref{fig:corner} shows two examples of one- and two-dimensional posterior distributions (cornerplots) of the model free parameters when fitting the RGB LF in the JKC passband for the SMC (left) and the LMC (right). In all other cases, the cornerplots are very similar to those shown in Fig.~\ref{fig:corner}, and we do not show them here for the sake of brevity. All parameters are well-constrained, with the RGB tip magnitude $\mtip$ exhibiting weak correlation with the other parameters. The most correlated parameters are always the slope $b$ and the discontinuity jump $c$. This dependence is easily understood, as a slight change in slope can compensate for a narrower/broader magnitude discontinuity when convolved by $\sigtip$. For all passbands and both galaxies, in Table~\ref{tab:params} we list the model median parameters alongside the $1\sigma$ confidence intervals resulting from the fitting procedure (see also Fig.s~\ref{fig:rem1} and \ref{fig:rem2}). 

\begin{figure}
        \centering
        \includegraphics[width=1\hsize]{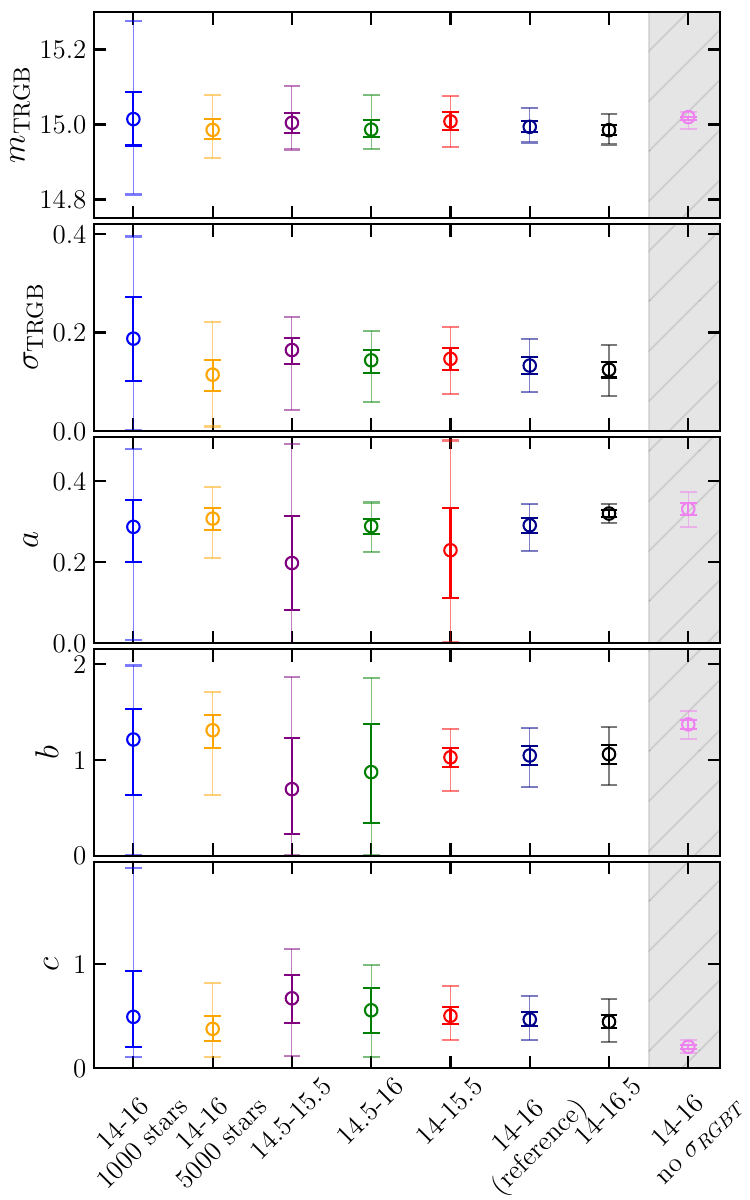}
    \caption{{ Tests performed on SMC sample in the JKC band (see Appendix~\ref{app:corner}). The panels, from top to bottom, show the inferred models parameters $\mtip$, $\sigtip$, $a$, $b$ and $c$, alongide the $1\sigma$ and $3\sigma$ confidence intervals in the cases considered.}}\label{fig:tests}
\end{figure}

{ We tested the robustness of the tip estimate against variations of the adopted portion of the upper RGB LF included in the fit, against the number of stars in the sample for a fixed magnitude range, and effect of introducing $\sigtip$ over a model with $\sigtip=0$. All tests were performed using the SMC sample in the JKC photometric system, though we expect similar trends in the LMC and in different photometric bands.}

{ Specifically, the tests considered are as follows:
\begin{itemize}
    \item[i)] 1000 stars between 14 and 16 mag;
    \item[ii)] 5000 stars between 14 and 16 mag;
    \item[iii)] All stars between 14.5 and 15.5 mag;
    \item[iv)] All stars between 14.5 and 16 mag;
    \item[v)] All stars between 14 and 15.5 mag;
    \item[vi)] All stars between 14 and 16 mag (the reference model);
    \item[vii)] All stars between 14 and 16.5 mag;
    \item[viii)] All stars between 14 and 16 mag but with $\sigma_{TRGB}=0$.
\end{itemize}
Figure~\ref{fig:tests} and Table \ref{tab:prmtest} schematically present the outcomes of our analysis, displaying the inferred values of the model parameters for all the considered cases. The errors are computed consistently with our definition. Below, we provide a few key observations: 

\begin{itemize}
    \item {\em Effect of sample size:} as it is reasonable to expect, using fewer stars within a fixed magnitude range results in larger errors in the model parameters (e.g., blue and orange points to the left in Figure~\ref{fig:tests}).
    \item {\em effect of magnitude range:} considering all stars within $\pm0.5$ mag from the tip (i.e., a magnitude range of 14.5-15.5) results in the most uncertain estimates for parameters $a$, $b$, and $c$. The estimate for $a$ improves when the magnitude range is increased towards higher magnitudes (14.5-16), while the estimate for $b$ improves when the magnitude range is extended towards lower magnitudes (14-15.5). Both estimates improve simultaneously when the magnitude range is 14-16. This result is expected since $a$ is sensitive to the less luminous end and $b$ to the more luminous end of the RGB LF, necessitating adequate sampling of these regions for accurate parameter constraints. When considering even wider magnitude ranges (14-16.5), the inference for the slope $a$ continues to improve, but there is no significant gain for the other parameters. Notably, the median tip magnitude and its corresponding errors remain approximately the same. This indicates that fitting broader portions of the LF does not improve the estimate of the tip. The precision of the tip estimate depends primarily on how well the tip itself is sampled, rather than the width of the LF included in the fit.
    \item {\em consistency of tip measurements:} in all cases considered, the tip measurements are consistent within the errors, despite variations in the estimates of other parameters.
    \item {\em effects of $\sigtip$:} the last case is intended to  test whether the introduction of $\sigma_{RGBT}$ introduces biases in the estimate of the tip magnitude. This is done by consistently fitting the same reference case (SMC in the JKC passband) with a $\sigma_{RGBT}=0$ model (pink points to the right of Figure \ref{fig:tests}). All parameters, except for $\mtip$, show variations that are, within the errors, poorly consistent with the corresponding model with $\sigtip$. On the other hand, the inferred tip magnitude remains within $<1.6\sigma$ of our reference value obtained with a model that includes $\sigtip$, showing that this parameter does not bias the measure.
    We note that introducing $\sigtip$ to the model yields a more realistic errors representation on $\mtip$ and larger errors on all other parameters. When $\sigma_{RGBT}=0$, the error on $\mtip$ shrinks to the order of the photometric error (0.005 mag), hence it is clearly underestimated. Although this could be interpreted as a very precise measurement of the tip, a close look to a comparison between data and model is enough to notice that this model gives an erroneous representation of the RGB LF. This lack of flexibility is the reason to the unrealistically small errors on the tip: when the LF presents a smooth transition and a broad RGB tip, a model with $\sigtip=0$ is simply inadequate 
    to accurately represent the data. Fig~\ref{fig:tipnposgima} shows the SMC RGB LF in the JKC alongside the median model and the $1-3\sigma$ confidence interval of the $\sigtip=0$ case. Due to this lack of flexibility, the model is strongly constrained around values that are not representative of the LF and is incapable of exploring better-fitting models, resulting in a statistically precise measurement, but inadequate error estimation.    
\end{itemize}

\begin{figure}
    \centering
    \includegraphics[width=1\hsize]{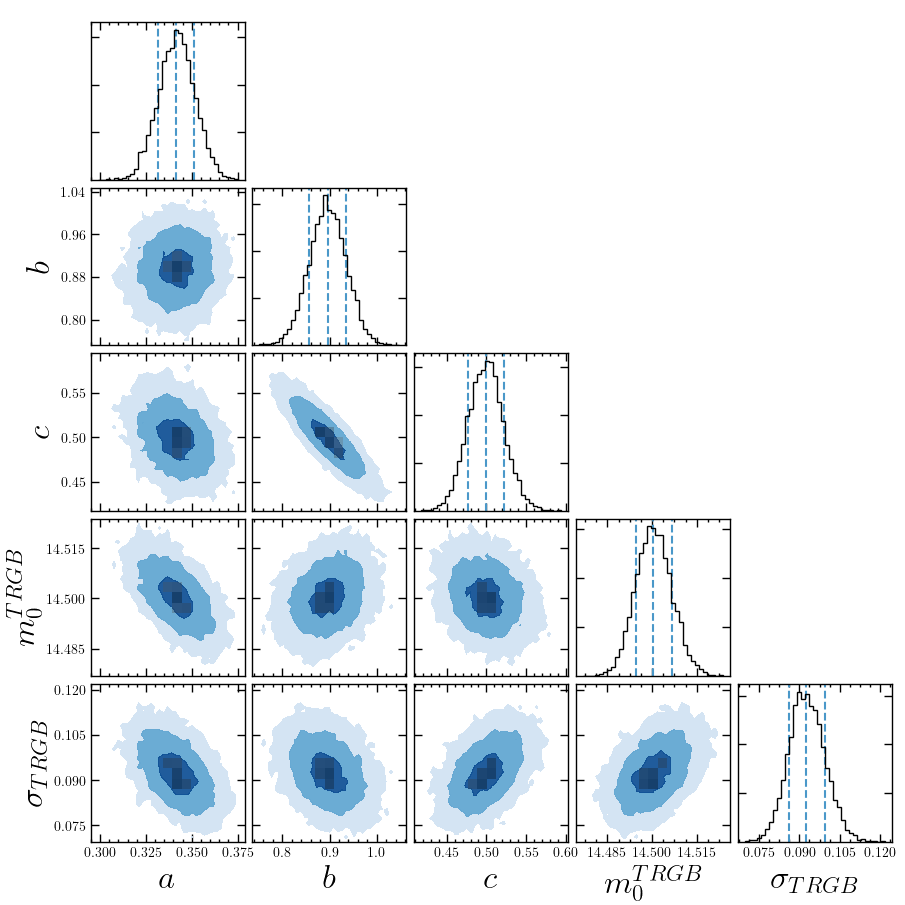}
    \caption{{ Detection of the RGB tip in the JKC band for the SMC fitting a model with $\sigtip=0$. The value of the model parameters and errors are listed to the bottom of Table~\ref{tab:prmtest}.}}    \label{fig:tipnposgima}
\end{figure}

Our choice to standardize to a range of 1.5 to 2 magnitudes around the tip is motivated by model convergence and the need to ensure adequate sampling of the RGB LF. We avoided considering larger magnitude ranges (such as 14 to 16.5 in the SMC) to reduce computational time. In the case of the LMC, a 2.5 magnitude range results in a significantly high number of stars, which sensibly slows down the fitting process without improving the accuracy of the tip estimate.}

\begin{table*}[h!]
    \centering
    \renewcommand{\arraystretch}{1.5}
    \begin{tabular}{cccccc}
        \hline
        test type & $a$ & $b$ & $c$ & $\mtip$ & $\sigtip$ \\
        \hline
        $[14,16]$, 1000 stars & $0.288_{-0.088(-0.279)}^{+0.067(+0.192)}$ & $1.210_{-0.574(-1.200)}^{+0.314(+0.772)}$ & $0.491_{-0.287(-0.390)}^{+0.438(+1.432)}$ & $15.014_{-0.070(-0.201)}^{+0.072(+0.261)}$ & $0.187_{-0.086(-0.187)}^{+0.085(+0.208)}$ \\
        $[14,16]$, 5000 stars & $0.308_{-0.029(-0.097)}^{+0.027(+0.078)}$ & $1.306_{-0.181(-0.679)}^{+0.159(+0.398)}$ & $0.376_{-0.117(-0.271)}^{+0.120(+0.439)}$ & $14.986_{-0.025(-0.074)}^{+0.028(+0.092)}$ & $0.114_{-0.033(-0.106)}^{+0.029(+0.107)}$ \\
        $[14.5,15.5]$  & $0.198_{-0.116(-0.197)}^{+0.117(+0.295)}$ & $0.692_{-0.466(-0.689)}^{+0.529(+1.166)}$ & $0.669_{-0.241(-0.556)}^{+0.221(+0.472)}$ & $15.004_{-0.026(-0.071)}^{+0.027(+0.098)}$ & $0.164_{-0.028(-0.121)}^{+0.025(+0.067)}$ \\
        $[14.5,16]$ & $0.289_{-0.020(-0.064)}^{+0.019(+0.059)}$ & $0.870_{-0.530(-0.866)}^{+0.497(+0.981)}$ & $0.554_{-0.215(-0.444)}^{+0.218(+0.433)}$ & $14.986_{-0.020(-0.051)}^{+0.026(+0.091)}$ & $0.143_{-0.026(-0.086)}^{+0.021(+0.059)}$ \\
        $[14,15.5]$ & $0.230_{-0.117(-0.228)}^{+0.105(+0.271)}$ & $1.024_{-0.104(-0.349)}^{+0.094(+0.295)}$ & $0.501_{-0.078(-0.228)}^{+0.084(+0.288)}$ & $15.008_{-0.024(-0.068)}^{+0.025(+0.067)}$ & $0.146_{-0.022(-0.072)}^{+0.022(+0.066)}$ \\
        $[14,16]$ & $0.291_{-0.019(-0.062)}^{+0.018(+0.052)}$ & $1.043_{-0.103(-0.332)}^{+0.096(+0.284)}$ & $0.466_{-0.065(-0.199)}^{+0.069(+0.229)}$ & $14.994_{-0.014(-0.042)}^{+0.015(+0.051)}$ & $0.133_{-0.016(-0.054)}^{+0.017(+0.054)}$ \\
        $[14,16.5]$ & $0.320_{-0.008(-0.023)}^{+0.008(+0.024)}$ & $1.057_{-0.099(-0.325)}^{+0.094(+0.281)}$ & $0.444_{-0.060(-0.190)}^{+0.065(+0.218)}$ & $14.985_{-0.013(-0.038)}^{+0.014(+0.044)}$ & $0.124_{-0.016(-0.054)}^{+0.015(+0.051)}$ \\
        $[14,16]$, $\sigtip=0$ & $0.332_{-0.015(-0.045)}^{+0.014(+0.042)}$ & $1.367_{-0.047(-0.152)}^{+0.047(+0.136)}$ & $0.203_{-0.021(-0.060)}^{+0.021(+0.066)}$ & $15.020_{-0.008(-0.032)}^{+0.000(+0.014)}$ & - \\
        \hline
    \end{tabular}
    \caption{Results of the tests performed on the SMC sample in the JKC band (see Appendix~\ref{app:corner} for details). The table reports, from left to right, the $1\sigma$ and $3\sigma$ values of $\mtip$, $\sigtip$, $a$, $b$, and $c$, while from top to bottom, lists the different tests.}
    \label{tab:prmtest}
\end{table*}


\begin{table*}[!htbp]
\label{tab:fitpar}
\renewcommand{\arraystretch}{1.5}
\begin{tabular}{lccccccc}
\hline
gal & phot. sys.  & mag  & $a$ & $b$ & $c$ & $\mtip$ & $\sigtip$ \\  
\hline  

SMC &   JKC      &  I	 & $0.291_{-0.019}^{+0.018}$ & $1.043_{-0.103}^{+0.096}$ & $0.466_{-0.065}^{+0.069}$ & $14.994_{-0.014}^{+0.015}$ & $0.133_{-0.016}^{+0.017}$ \\ 
LMC &   JKC      &  I	 & $0.342\pm0.010$ & $0.896\pm0.039$ & $0.500_{-0.023}^{+0.022}$ & $14.500\pm0.006$ & $0.092_{-0.006}^{+0.007}$  \\ 
SMC & ACS-WFC    &F814W  & $0.273_{-0.020}^{+0.018}$ & $0.959_{-0.112}^{+0.110}$ & $0.542_{-0.072}^{+0.076}$ & $14.981\pm0.014$ & $0.143\pm0.016$ \\ 
LMC & ACS-WFC    &F814W  & $0.318\pm0.010$ & $0.951_{-0.044}^{+0.042}$ & $0.530_{-0.025}^{+0.027}$ & $14.493\pm0.006$ & $0.102\pm0.007$ \\ 
SMC &  SDSS	 &  i	 & $0.267_{-0.048}^{+0.043}$ & $0.876_{-0.172}^{+0.149}$ & $0.580_{-0.088}^{+0.102}$ & $15.497_{-0.016}^{+0.017}$ & $0.141_{-0.017}^{+0.019}$  \\ 
LMC &  SDSS	 &  i	 & $0.318_{-0.022}^{+0.021}$ & $0.850_{-0.047}^{+0.044}$ & $0.475_{-0.024}^{+0.025}$ & $15.023_{-0.006}^{+0.007}$ & $0.086\pm0.007$ \\ 
SMC &  SDSS	 &  z	 & $0.250_{-0.084}^{+0.070}$ & $0.690_{-0.120}^{+0.115}$ & $0.623_{-0.078}^{+0.085}$ & $15.191_{-0.022}^{+0.024}$ & $0.149_{-0.018}^{+0.021}$  \\ 
LMC &  SDSS	 &  z	 & $0.372_{-0.024}^{+0.022}$ & $0.368\pm0.049$ & $0.607_{-0.029}^{+0.030}$ & $14.621\pm0.007$ & $0.097_{-0.007}^{+0.008}$  \\ 
SMC &	PS1	 &  y	 & $0.268_{-0.056}^{+0.048}$ & $0.575_{-0.142}^{+0.130}$ & $0.630_{-0.078}^{+0.087}$ & $15.099_{-0.017}^{+0.019}$ & $0.141_{-0.018}^{+0.019}$ \\ 
LMC &	PS1	 &  y	 & $0.309_{-0.021}^{+0.020}$ & $0.282_{-0.075}^{+0.070}$ & $0.695_{-0.039}^{+0.041}$ & $14.530_{-0.007}^{+0.008}$ & $0.116_{-0.008}^{+0.009}$ \\ 
SMC &JWST-NIRCAM & F090W & $0.239_{-0.036}^{+0.033}$ & $1.005\pm0.063$ & $0.424_{-0.045}^{+0.048}$ & $14.717_{-0.015}^{+0.017}$ & $0.119\pm0.016$  \\ 
LMC &JWST-NIRCAM & F090W & $0.308_{-0.017}^{+0.016}$ & $0.882_{-0.035}^{+0.034}$ & $0.506_{-0.023}^{+0.024}$ & $14.169\pm0.008$ & $0.107\pm0.008$ \\
SMC &  NGRST	 & Z087  & $0.229_{-0.059}^{+0.053}$ & $0.905_{-0.097}^{+0.096}$ & $0.665_{-0.081}^{+0.086}$ & $14.783_{-0.020}^{+0.022}$ & $0.175_{-0.019}^{+0.020}$ \\ 
LMC &  NGRST	 & Z087  & $0.346_{-0.020}^{+0.019}$ & $1.034\pm0.034$ & $0.472\pm0.024$ & $14.259\pm0.008$ & $0.101\pm0.008$  \\ 
SMC &  Euclid-VIS    & I$_E$   & $0.288_{-0.022}^{+0.021}$ & $1.203_{-0.103}^{+0.096}$ & $0.450_{-0.069}^{+0.076}$ & $15.677_{-0.014}^{+0.015}$ & $0.136_{-0.018}^{+0.017}$  \\ 
LMC &  Euclid-VIS    & I$_E$   & $0.317_{-0.016}^{+0.015}$ & $0.907_{-0.065}^{+0.062}$ & $0.573_{-0.031}^{+0.032}$ & $15.230_{-0.005}^{+0.006}$ & $0.095_{-0.006}^{+0.007}$  \\ 
\hline
\end{tabular}
\caption{$1\sigma$ confidence intervals on the model free parameters as a result of the fitting algorithm of Section~\ref{sec:method}. From left to right: target galaxy (SMC or LMC), photometric system, magnitude, $a$, $b$, $c$, $\mtip$ and $\sigtip$ from model ~\ref{for:final}.}\label{tab:params}
\end{table*}

\end{appendix}

\end{document}